\documentclass[a4paper,journal]{IEEEtran}

\usepackage[utf8]{inputenc}
\usepackage{graphicx}
\usepackage{float}
\usepackage{color, colortbl}
\usepackage{xcolor}
\usepackage{array}
\usepackage{multirow}
\usepackage{footnote}
\usepackage{cite}
\usepackage{lipsum}
\usepackage{threeparttable}
\usepackage{amsmath}
\usepackage{mathtools}
\usepackage[linesnumbered,vlined]{algorithm2e}

\usepackage[hyphens]{url}

\usepackage{soul}

\makeatletter
\def\blfootnote{\xdef\@thefnmark{}\@footnotetext}
\makeatother

\usepackage{lipsum}

\begin{document}
 \title{A High Efficiency MAC Protocol for WLANs: Providing Fairness in Dense Scenarios}

  \author{
      \IEEEauthorblockN{Luis Sanabria-Russo\IEEEauthorrefmark{0}, Jaume Barcelo\IEEEauthorrefmark{0}, Boris Bellalta\IEEEauthorrefmark{0}, Francesco Gringoli\IEEEauthorrefmark{0}}}



\maketitle

\blfootnote{This work was partially supported by the Spanish government, through the project CISNETS (TEC2012-32354).\\
L. Sanabria-Russo (luis.sanabria@upf.edu) , J. Barcelo (jaume.barcelo@upf.edu) and B. Bellalta (boris.bellalta@upf.edu) are with the Department of Information and Communication Technologies, Universitat Pompeu Fabra, Barcelona, Spain.\\ F. Gringoli (francesco.gringoli@ing.unibs.it) is with the Dipartamento di Elettronica per l’Automazione, Universit\`{a} degli Studi di Brescia, Brescia, Italy.}

\begin{abstract}

\boldmath Collisions are a main cause of throughput degradation in WLANs. The current contention mechanism used in IEEE 802.11 networks is called Carrier Sense Multiple Access with Collision Avoidance (CSMA/CA). It uses a Binary Exponential Backoff (BEB) technique to randomise each contender attempt of transmitting, effectively reducing the collision probability. Nevertheless, CSMA/CA relies on a random backoff that while effective and fully decentralised, in principle is unable to completely eliminate collisions, therefore degrading the network throughput as more contenders attempt to share the channel. 

To overcome these situations, Carrier Sense Multiple Access with Enhanced Collision Avoidance (CSMA/ECA) is able to create a collision-free schedule in a fully decentralised manner using a deterministic backoff after successful transmissions. Hysteresis and Fair Share are two extensions of CSMA/ECA to support a large number of contenders in a collision-free schedule. CSMA/ECA offers better throughput than CSMA/CA and short-term throughput fairness. This work describes CSMA/ECA and its extensions. Additionally, it provides the first evaluation results of CSMA/ECA with non-saturated traffic, channel errors, and its performance when coexisting with CSMA/CA nodes. Furthermore, it describes the effects of imperfect clocks over CSMA/ECA and present a mechanism to leverage the impact of channel errors and the addition/withdrawal of nodes over collision-free schedules. Finally, experimental results on throughput and lost frames from a CSMA/ECA implementation using commercial hardware and open-source firmware are presented.
\end{abstract}


\section{Introduction}\label{introduction}
Wireless Local Area Networks (WLANs or WiFi networks~\cite{802Standards}) are a popular solution for wireless connectivity, whether in public places, work environments or at home. This technology works over an unlicensed spectrum in the Industrial, Scientific and Medical (ISM) radio bands (at around $2.4$ or $5$~GHz), offering a good tradeoff between performance and costs, which is a main reason for its popularity. 

The Medium Access Control (MAC) scheme used in WLANs is based on Carrier Sense Multiple Access with Collision Avoidance (CSMA/CA) protocol. It has been widely adopted by manufacturers and consumers, making it inexpensive to implement and an ubiquitous technology. Nevertheless, the ever-growing throughput demands from upper layers are faced with a bottleneck at the WLANs' MAC~\cite{perahia2008ieee}, which by its nature is prone to collisions that degrade the overall performance as more nodes join the network~\cite{bianchi2000performance}.

The research community has pushed forward many alternatives to the current MAC in WLANs~\cite{bharghavan1994map,wang2004ncr,cali2000dti,lopez-toledo2006aoi,
barcelo2008lba,bellalta2009vtc,HE,CSMA_ECA,L_MAC2,hui2011epp,barcelo2011tcf}, but when a proposal deviates too much from CSMA/CA, or some time-critical operations are modified, its hardware implementation as part of WLANs' MAC often becomes unlikely~\cite{WMP}, with the standardisation process taking many years without certainty of approval~\cite{perahia2008ieee}. 

A CSMA/CA replacement should be able to provide advantages in terms of throughput, spectrum efficiency and number of supported contenders. All of the aforementioned without sacrificing short-term throughput fairness. Furthermore, WLANs implementing such a replacement must also serve existing users, which means they have to be backwards compatible.

A suitable candidate, and the one to be evaluated in this work, is called Carrier Sense Multiple Access with Enhanced Collision Avoidance (CSMA/ECA)~\cite{barcelo2008lba}. It is capable of attaining higher throughput than CSMA/CA by making a simple modification to the contention mechanism, thus keeping maintaining compatibility. In CSMA/ECA, nodes use a deterministic backoff after successful transmissions, constructing a collision-free schedule among successful contenders in a fully decentralised way. This backoff mechanism ensures that more channel time is spent on successful transmissions rather than recovering from collisions, thus increasing the throughput of the network. Further enhancements (or extensions), like \emph{Hysteresis} and \emph{Fair Share}~\cite{research2standards} allow CSMA/ECA to support many more contenders in a collision-free schedule. Moreover, CSMA/ECA and its extensions are designed for allowing nodes to transmit as frequently as possible while keeping an even distribution of the available bandwidth among users.

The 802.11 High Efficiency WLAN (HEW) Task Group (TG) envisions very crowded scenarios as one of the future challenges for WLAN protocols~\cite{HEW-scenarios,bellalta2015WCM}, specifically those usually encountered at stadiums or conference rooms. Further, if the need for serving many users is combined with the increased throughput demands from the upper layers, a performance improvement at the MAC becomes paramount. Although many studies have been made analysing the performance of CSMA/ECA~\cite{barcelo2008lba,research2standards,bellalta2009vtc,E2CA_performance}, there are still several open aspects that require further attention and additional insight to consider CSMA/ECA as a potential CSMA/CA replacement. Namely, neither assesses the protocol's backwards compatibility property or its behavior under non-saturated traffic conditions while serving many users. Furthermore, the impact of channel errors, node addition/withdrawal, and imperfect clocks over the deterministic backoff mechanism is also lacking.

This paper fills those gaps by extending~\cite{research2standards}, and consolidates the push for CSMA/ECA as a potential replacement of CSMA/CA for next generation WLANs. In detail, this paper provides the following contributions:

\begin{itemize}
	\item First results on the achievable throughput and delay of CSMA/ECA with Hysteresis and Fair Share under non-saturated traffic conditions for very large number of nodes.
	\item The impact of imperfect clocks and channel errors on CSMA/ECA with Hysteresis and Fair Share's deterministic backoff and its consequences on the achieved performance.
	\item Formulation of the throughput bounds for CSMA/ECA with Hysteresis and Fair Share.
	\item Introduce the Schedule Reset mechanism for reducing the schedule length in case of users withdrawing from the contention (non-saturated traffic), or when its length is increased due to channel errors.
	\item Coexistence and backwards compatibility with CSMA/CA nodes under different traffic conditions.
	\item First implementation of CSMA/ECA with Hysteresis in real hardware and the experimental results in a crowded WLAN testbed.
\end{itemize}

Results, derived from a modified version of the COST~\cite{COST} simulator show that CSMA/ECA with Hysteresis and Fair Share is capable of accommodating many users in collision-free schedules. As tests in a mixed network with different proportions of CSMA/CA and CSMA/ECA nodes show, the aggregated throughput is higher than the observed in CSMA/CA-only networks. Furthermore, at low number of total contenders CSMA/CA's throughput is actually improved, while it is degraded in crowded scenarios. This constitutes a motivation for a change towards CSMA/ECA.

Beyond simulations results, the implementation of CSMA/ECA prototypes\cite{ECA-DEMO-INFOCOM14, sanabria2013prototyping, BECA-test,CF-MAC} show that the construction of collision-free schedules using a deterministic backoff after successful transmissions is possible and results in a throughput increase by reducing the number of corrupted frames. We then present the first real hardware implementation of CSMA/ECA with Hysteresis using the open firmware OpenFWWF~\cite{OpenFWWF}. Results show that CSMA/ECA is able of providing higher throughput and lower losses than CSMA/CA for the same number of contenders.

An overview of similar decentralised and collision-free MAC protocols for WLANs is provided in Section~\ref{relatedWork}. CSMA/ECA, as well as its extensions for allocating many contenders in a collision-free schedule are explained in Section~\ref{introProtocol}. Section~\ref{simulations} details the simulation environment for testing CSMA/ECA, while Section~\ref{results} explains the results. An overview of CSMA/ECA real-hardware implementation and testbed description is compiled in Section~\ref{EDCA}, followed by a summary of the still missing CSMA/ECA features needed to become the next CSMA/CA replacement in Section~\ref{ECAtoCA}. Conclusions are drawn in Section~\ref{conclusions}.

\section{Related Work}\label{relatedWork}
Time in WLANs is divided into tiny empty slots of fixed length $\sigma_{e}$, collisions, and successful slots of length $\sigma_{c}$ and $\sigma_{s}$, respectively. Collision and successful slots contain collisions or successful transmissions, making them several orders of magnitude larger than empty slots ($\sigma_{e}\ll\min(\sigma_{s},\sigma_{c}))$. One of the effects of collisions is the degradation of the network performance by wasting channel time on collisions slots. 

Recent advances in the WLANs PHY~\cite{perahia2008ieee,6191306} push the research community towards the development of MAC protocols able to take advantage of a much faster PHY. By reducing the time spent in collisions nodes are able to transmit more often, which in turn translates to an increase in the network throughput. Further, the upcoming MAC protocols for WLANs should work without message exchange between contenders, that is, work in a fully decentralised fashion in order to avoid injecting extra control traffic that may reduce the data throughput.

Performing time slot reservation for each transmission is a well known technique for increasing the throughput and mantanining Quality of Service (QoS) in TDMA schemes, like LTE~\cite{canoLTEcoexistence}. Applying the same concept to CSMA networks by modifying DCF's random backoff proceedure provides similar benefits~\cite{HE}. The following are MAC protocols for WLANs, decentralised and capable of attaining greater throughput than CSMA/CA by constructing collision-free schedules using reservation techniques. A survey of collision-free MAC protocols for WLANs is presented in~\cite{L_MAC}. In this paper we only overview those that are similar to CSMA/ECA.

\subsection{Zero Collision MAC}\label{ZC-MAC}

Zero Collision MAC (ZC-MAC)~\cite{ZMAC} achieves a zero collision schedule for WLANs in a fully decentralised way. It does so by allowing contenders to reserve one empty slot from a predefined virtual schedule of $M$-slots in length. Backlogged stations pick a slot in the virtual cycle to attempt transmission. If two or more stations picked the same slot in the cycle, their transmissions will eventually collide. This forces the involved contenders to randomly and uniformly select other empty slot from those detected empty in the previous cycle plus the slot where they collided. When all $N$ stations reserve a different slot, a collision-free schedule is achieved.

ZC-MAC is able to outperform CSMA/CA under different scenarios. Nevertheless, given that the length of ZC MAC's virtual cycle has to be predefined without actual knowledge of the real number of contenders in the deployment, the protocol is unable to provide a collision-free schedule when $N>M$. Furthermore, if $M$ is overestimated ($M\gg N$), the fixed-width empty slots between each contender's successful transmission are no longer negligible and contribute to the degradation of the network performance. Additionally, ZC-MAC nodes require common knowledge of where the virtual schedule starts/ends. This is not considered in CSMA/CA and constitutes an obstacle towards standardisation.

\subsection{Learning-MAC}

Learning-MAC~\cite{L_MAC} is another MAC protocol able to build a collision-free schedule for many contenders. It does so defining a \emph{learning strength} parameter, $\beta\in(0,1)$. Each contender starts by picking a slot $s$ for transmission of the schedule $n$ of length $C$ at random with uniform probability. After a contender picks slot $s(n)$, its selection in the next schedule, $s(n+1)$, will be conditioned by the result of the current attempt. (\ref{success}) and (\ref{collisions-eq}) extracted from~\cite{L_MAC} show the probability of selecting the same slot $s(n)$ in cycle $n+1$.

\begin{equation} \label{success}
		\left. \begin{aligned}
			p_{s(n)}(n+1)&=1,\\
			p_{j}(n+1)&=0,
		\end{aligned}
	\right\}
	\qquad \text{\emph{Success}}
\end{equation}
\begin{equation} \label{collisions-eq}
	\left. \begin{aligned}
			p_{s(n)}(n+1)&=\beta p_{s(n)}(n),\\
			p_{j}(n+1)&=\beta p_{j}(n)+\frac{1-\beta}{C -1},
		\end{aligned}
	\right\}
	\qquad \text{\emph{Collision}}
\end{equation}
\\
for all $j\neq s(n),~j\in \{1,\dots ,C\}$. That is, if a station successfully transmitted in $s(n)$, it will pick the same slot on the next schedule with probability one. Otherwise, it follows~(\ref{collisions-eq}).

The selection of $\beta$ implies a compromise between fairness and convergence speed, which the authors determined $\beta=0.95$ to provide satisfactory results.

L-MAC is able to achieve higher throughput than CSMA/CA with a very fast convergence speed. Nevertheless, the choice of $\beta$ suppose a previous knowledge of the number of empty slots ($C-N$, where $N$ is the number of contenders), which is not easily available to CSMA/CA or may require a centralised entity~\cite{barcelo2011tcf}.

Further extensions to L-MAC introduced an \emph{Adaptative} schedule length in order to increase the number of supported contenders in a collision-free schedule. This adaptive schedule length is doubled or halved depending on the presence of collisions or many empty slots per schedule, respectively. The effects of reducing the schedule length may provoke a re-convergence phase which can result in short-term fairness issues. Moreover, L-MAC nodes also require common knowledge of the start/end of the schedule.\section{Carrier Sense Multiple Access with Enhanced Collision Avoidance (CSMA/ECA)}\label{introProtocol}

CSMA/ECA~\cite{barcelo2008lba} is a fully decentralised and collision-free MAC for WLANs. It differs from CSMA/CA in that it uses a deterministic backoff, $B_{\text{d}}=\lceil CW_{\min}/2\rceil-1$ after successful transmissions, where $CW_{\min}$ is the minimum contention window of typical value $CW_{\min}=16$. By doing so, contenders that successfully transmitted on schedule $n$, will transmit without colliding with other successful nodes in future cycles~\cite{HE}.

Collisions are handled as in CSMA/CA, which is described in Algorithm~\ref{alg:CSMA_CA}. In Algorithm~\ref{alg:CSMA_CA}, the node starts by setting the retransmissions counter and backoff stage to zero ($r\in[0,R]$ and $k\in[0,m]$ respectively, with $m$ the maximum backoff stage, and $R=m+1$ the retransmissions limit. The typical value for $m$ is $5$), then generates a random backoff, $B$. When the Acknowledgement (\emph{ack}) for a sent packet is not received by the sender a collision is assumed. Upon collision, the involved nodes will double their contention window by incrementing their backoff stage in one and use a random backoff, $B\in[0,2^{k}CW_{\min}-1]$. This procedure is described between Line~\ref{collision} and~\ref{finalCollision} of Algorithm~\ref{alg:CSMA_CA}.

\begin{figure*}[tb]
\centering
  \includegraphics[width=0.8\linewidth]{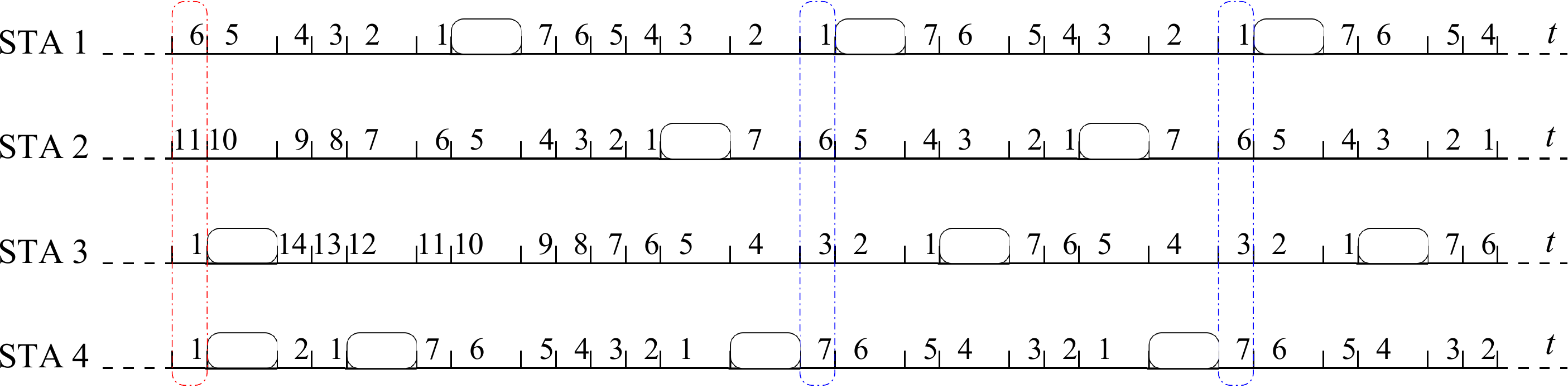}
  \caption{An example of the temporal evolution of CSMA/ECA in saturation. Slots containing transmissions are around 1000 times larger than empty slots}
  \label{fig:BECA-example}
\end{figure*}

Algorithm~\ref{alg:CSMA_ECA} provides an explanation of CSMA/ECA's deterministic backoff mechanism, which main difference with CSMA/CA (and therefore with Algorithm~\ref{alg:CSMA_CA}) relies on the selection of a deterministic backoff after a successful transmission (compare Line~\ref{randomBackoff} in Algorithm~\ref{alg:CSMA_CA} with Line~\ref{deterministicBackoff} in Algorithm~\ref{alg:CSMA_ECA}). Figure~\ref{fig:BECA-example} shows an example of CSMA/ECA dynamics with four contenders.

\begin{algorithm}[ht!!!]
\While{the device is on}
{
  $r \leftarrow 0$; $k \leftarrow 0$\;
  $B \leftarrow \mathcal{U}[0,2^k{\rm{CW}_{min}}-1]$\;
  \While{there is a packet to transmit}{
    \Repeat{($r = R$) or (success)}{
      \While{$B>0$}{
        wait 1 slot\;
        $B \leftarrow B-1$\;
      }
      \colorbox{yellow}{Attempt transmission of 1 packet;}\\
      \If{collision}{\label{collision}
        $r \leftarrow r+1$\;
        $k \leftarrow \min (k+1,m)$\;
        $B \leftarrow \mathcal{U}[0, 2^k {\rm{CW}_{min}} -1]$\;\label{finalCollision}
      }
    }
    $r \leftarrow 0$\;
    \colorbox{yellow}{$k \leftarrow 0$;}\\
	 \eIf{success}{
      \colorbox{yellow}{$B \leftarrow \mathcal{U}[0,2^{k}{\rm{CW}_{min}}-1]$;}\\\label{randomBackoff}
    }
    {
      Discard packet\;
      $B \leftarrow \mathcal{U}[0,2^k {\rm{CW}_{min}}-1]$\;
    }
  }
  Wait until there is a packet to transmit\;
}
\caption{\small{CSMA/CA. $r$ indicates the number of retransmission attempts, while $R$ is the maximum retransmission attempts limit. When $R$ retransmissions are reached, the packet waiting for transmission is dropped.}}
\label{alg:CSMA_CA}
\end{algorithm}

\begin{algorithm}[ht!!!]
\While{the device is on}
{
  $r \leftarrow 0$ ; $k \leftarrow 0$\;
  $B \leftarrow \mathcal{U}[0,2^k{\rm{CW}_{min}}-1]$\;
  \While{there is a packet to transmit}{
    \Repeat{($r = R$) or (success)}{
      \While{$B>0$}{
        wait 1 slot\;
        $B \leftarrow B-1$\;
      }
      \colorbox{yellow}{Attempt transmission of 1 packet;}\\
      \If{collision}
      {
        $r \leftarrow r+1$\;
        $k \leftarrow \min (k+1,m)$\;
        $B \leftarrow \mathcal{U}[0, 2^k {\rm{CW}_{min}} -1]$\;
      }
    }
    $r \leftarrow 0$\;
    \colorbox{yellow}{$k \leftarrow 0$;}\\
    \eIf{success}{
      \colorbox{yellow}{$B_{d} \leftarrow \lceil 2^{k}{\rm{CW}_{min}}/2\rceil-1$\;}\label{deterministicBackoff}\\
	 $B \leftarrow B_{d}$\;
    }
    {
      Discard packet\;
      $B \leftarrow \mathcal{U}[0, 2^k {\rm{CW}_{min}}-1]$\;
    }
  }
  Wait until there is a packet to transmit\;
}
\caption{\small{CSMA/ECA.}}
\label{alg:CSMA_ECA}
\end{algorithm}

In Figure~\ref{fig:BECA-example}, the \emph{STA-\#} labels represent stations willing to transmit. The horizontal lines represent a time axis with each number indicating the amount of empty slots left for the backoff to expire. Stations willing to transmit begin the contention for the channel by waiting a random backoff, $B$. The first outline highlights the fact that stations STA-3 and STA-4 will eventually collide because they have selected the same $B$. After recomputing the random backoff, STA-4's attempt results in a successful transmission, which instructs the node to use a deterministic backoff, $B_{\text{d}}=7$ in this case. By doing so, all successful STAs will not collide among each other in future cycles.

\begin{figure}[tb]
\centering
  \includegraphics[width=0.7\linewidth, angle=-90]{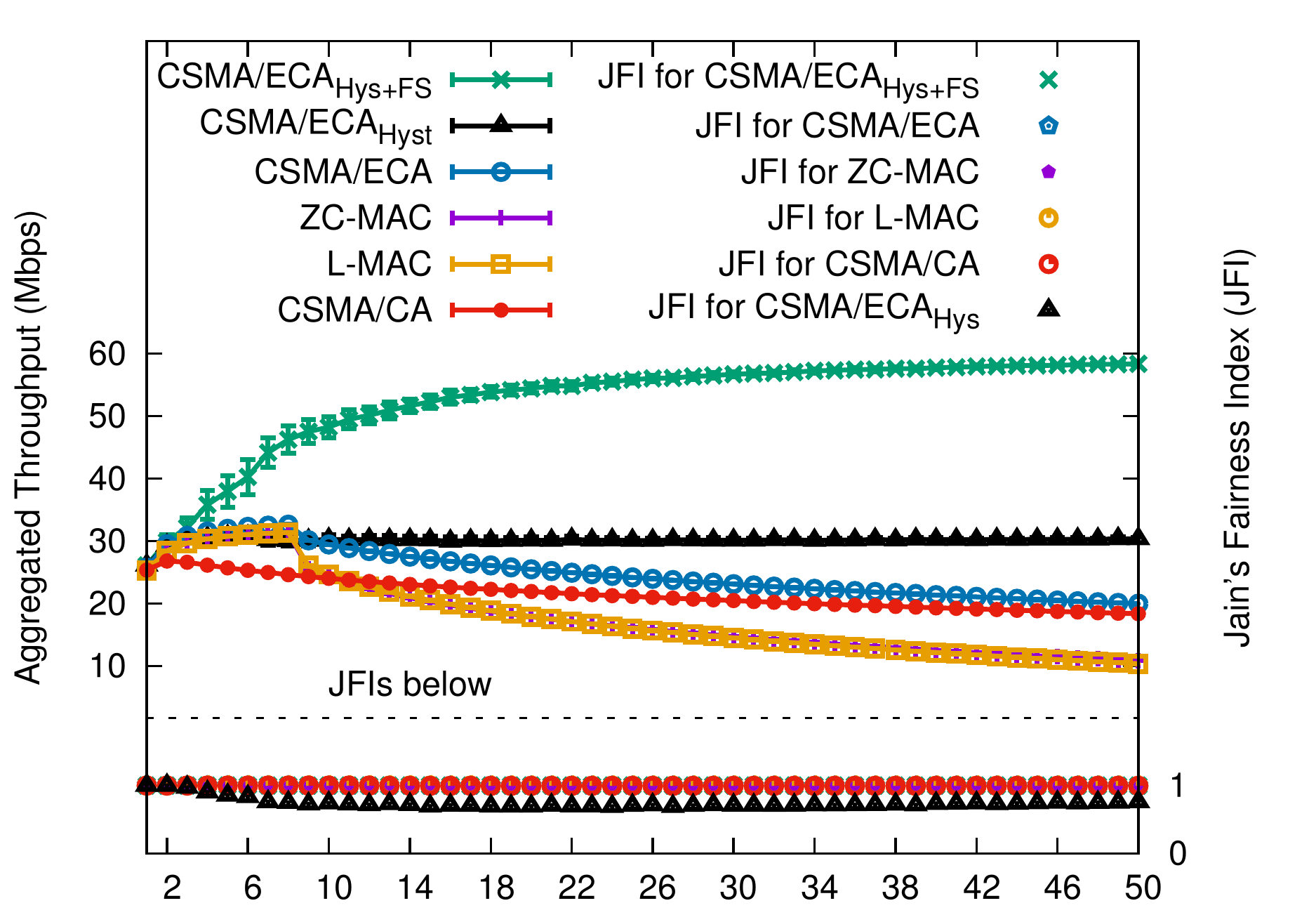}
  \caption{CSMA/ECA example in saturation: throughput (simulation parameters can be found in Section~\ref{simulations})}
  \label{fig:BECA}
\end{figure}

Collision slots being orders of magnitude larger than empty slots degrade the network performance. When CSMA/ECA builds the collision-free schedule all contenders are able to successfully transmit more often, increasing the aggregated throughput beyond CSMA/CA's. Figure~\ref{fig:BECA}, shows the average aggregated throughput of CSMA/ECA, CSMA/CA and collision-free protocolos presented in Section~\ref{relatedWork}, namely ZC-MAC and L-MAC~\cite{L_MAC}.


Referring to Figure~\ref{fig:BECA}, CSMA/ECA is able to achieve an aggregated throughput that goes beyond CSMA/CA up until the number of contenders is greater than $B_{\text{d}}+1$ ($N=8$ in the case of the figure). Beyond this point, the network will have a mixed behaviour relating to backoff mechanisms: some nodes will successfully transmit and use a deterministic backoff while others will collide due to the lack of empty slots and return to a random backoff. As more contenders join the network, CSMA/ECA performance will approximate to CSMA/CA's. This effect is also observed in ZC-MAC and L-MAC. It happens becasue the virtual cicle used by these protocolos ($M$ for ZC-MAC and $C$ for L-MAC in Section~\ref{relatedWork}) in lower than the number of contenders, $N$\footnote{L-MAC and ZC-MAC curves in Figure~\ref{fig:BECA} appear to yield the same aggregated throughput. This is because these protocols do not consider an increase in the backoff stage after a failed transmission, augmenting the collision probability beyond CSMA/CA's. An adaptation of the schedule length leverages this issue, which is preseted in~\cite{L_MAC} as \emph{Adaptive} L-MAC and L-ZC.}.

The \emph{JFI } curves in Figure~\ref{fig:BECA} show the Jain's Fairness index for all tested protocols. Showing a JFI equal to one suggests that the available throughput is shared evenly among all stations.

	\subsection{Supporting many more contenders}\label{moreContenders}
	As was mentioned before, CSMA/ECA is only able to build a collision-free schedule if the number of contenders $N$, is less or equal than $B_{\text{d}}+1$. When $N > B_{\text{d}}+1$, collisions reappear. 
	
	To be able to attain a collision-free schedule even when the number of contenders exceeds $B_{\text{d}}+1$, we introduce \emph{Hysteresis}. Hysteresis is a property of the protocol that instructs nodes not to reset their backoff stage ($k$) after successful transmissions, but to use a deterministic backoff $B_{\text{d}}=\lceil CW(k)/2\rceil -1$, where $CW(k)=2^{k}CW_{\min}$. This measure allows the adaptation of the schedule length, admitting many more contenders in a collision-free schedule. This idea of a schedule is significantly different from the virtual schedule required by the protocols described in Section~\ref{relatedWork}, that is, CSMA/ECA with Hysteresis does not require a previous knowledge of the number of contenders, the result of previous transmissions or the start/end of the schedule, easing its implementation in real hardware.
	
	Hysteresis enables CSMA/ECA nodes to have different schedules ($B_{\text{d}}$), carrying the undesired effect of unevenly dividing the channel time among contenders (i.e., some nodes will have to wait more in order to attempt transmissions).
	
	This unfairness issue is solved by instructing nodes at backoff stage $k$ to transmit $2^{k}$ packets on each attempt, thus proportionally compensating those nodes at higher backoff stages. This additional extension to CSMA/ECA is called \emph{Fair Share}. CSMA/ECA with Hysteresis and Fair Share will be referred to as CSMA/ECA$_{\text{Hys+FS}}$ in order to distinguish it from what was described until this point.
	
	The idea of allowing the transmission of more packets to stations that transmit less often was initially proposed by Fang et al. in~\cite{L_MAC}. It was later adapted to CSMA/ECA$_{\text{Hys}}$ and named Fair Share in~\cite{research2standards}. Figure~\ref{fig:BECA} shows the JFI for CSMA/CA as well as for CSMA/ECA$_{\text{Hys+FS}}$.
	
	
	In Figure~\ref{fig:BECA}, the only curve deviating from JFI = 1 is CSMA/ECA with Hysteresis (\emph{CSMA/ECA$_{\text{Hys}}$}), suggesting an uneven partition of the channel access time among contenders (which is fixed with Fair Share).
	
	\begin{figure*}[tb]
	\centering
		\includegraphics[width=0.8\linewidth]{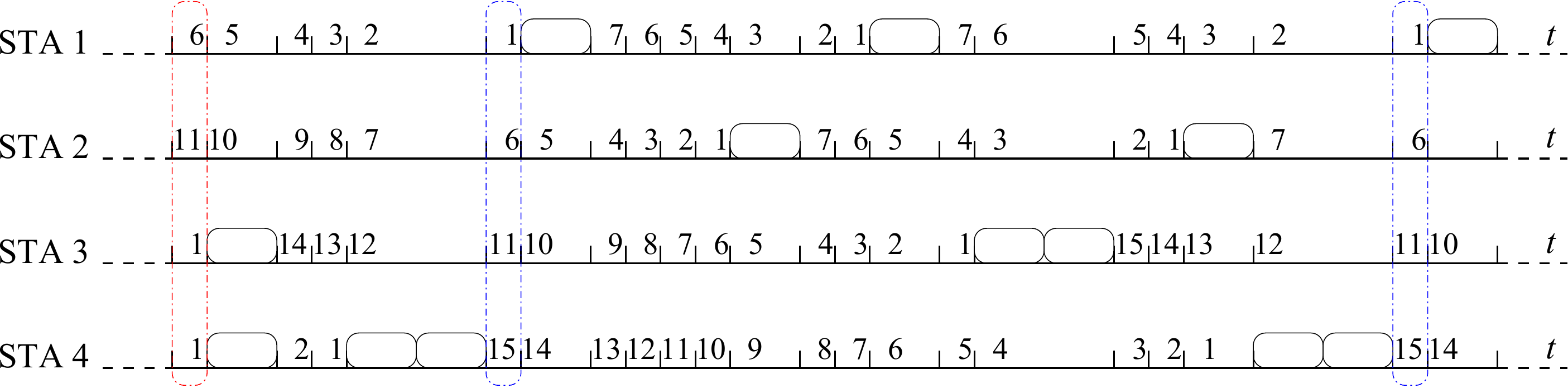}
		\caption{An example of the temporal evolution of CSMA/ECA$_{\text{Hys+FS}}$ in saturation ($CW_{\min}=16$)}
		\label{fig:ECA+Hyst}
	\end{figure*}
	
	\begin{table}
		\centering
		\caption{CSMA/ECA Notation}
		\label{tab:glossary}
		\begin{tabular}{|p{0.3\linewidth}|p{0.6\linewidth}|}
			\hline
			{\bfseries Notation} & {\bfseries Description}\\
			\hline
			$k$ & Backoff stage\\
			\hline
			$m$ & Maximum backoff stage\\
			\hline
			$B$ & Random backoff\\
			\hline
			$B_{\text{d}}$ & Deterministic backoff\\
			\hline
			$CW_{\min}$ & Minimum Contention Window\\
			\hline
			CSMA/ECA$_{\text{Hys}}$ & CSMA/ECA with Hysteresis\\
			\hline
			CSMA/ECA$_{\text{Hys+FS}}$ & CSMA/ECA with Hysteresis and Fair Share\\
			\hline
			CSMA/ECA$_{\text{Hys+MaxAg}}$ & CSMA/ECA with Hysteresis and Maximum Aggregation\\
			\hline
			CSMA/CA$_{\text{FS}}$ & CSMA/CA with Fair Share\\
			\hline
			CSMA/ECA$_{\text{MaxAg}}$ & CSMA/CA with Maximum Aggregation\\
			\hline
		\end{tabular}
	\end{table}

	Algorithm~\ref{alg:fullECA} describes CSMA/ECA$_{\text{Hys+FS}}$, Table~\ref{tab:glossary} provides a short list of notations used throughout the text, while an example of CSMA/ECA$_{\text{Hys+FS}}$ with four contenders is shown in  Figure~\ref{fig:ECA+Hyst}. In the figure the first outline indicates a collision between STA-3 and STA-4, which will provoke an increment on both stations' backoff stage ($k\leftarrow k+1$). Once STA-4's random backoff expires, CSMA/ECA$_{\text{Hys+FS}}$ instructs the station to transmit $2^{k}$ packets, and then use a deterministic backoff, $B_{\text{d}}=\lceil CW(k)/2\rceil-1$. The same procedure is followed by STA 3.
	
	With Hysteresis and Fair Share, CSMA/ECA$_{\text{Hys+FS}}$ is able to achieve greater throughput than CSMA/CA and for many more contenders, as shown also in Figure~\ref{fig:BECA}. In the figure, the \emph{CSMA/ECA$_{\text{Hys+FS}}$} curve shows a greater throughput because collisions are eliminated and Fair Share allows nodes to send $2^{k}$ packets upon each transmission. This throughput increase is the result of aggregation via Fair Share. It carries the negative effect of raising the average time between successful transmissions (see Section~\ref{timeBetweenSxTx}), which may affect delay-sensitive traffic, like gaming or live video/voice/tv streaming. 
	
	Channel errors, hidden-nodes, or unsaturated traffic conditions will disrupt any collision-free schedule, generate collisions and push all stations' backoff stages to the maximum value very quickly (contributing to the delay increase). This issue is leveraged with the Schedule Reset mechanim, introduced in Section~\ref{schedReset}.

	\begin{algorithm}[tb]
	\While{the device is on}
	{
	  $r \leftarrow 0$ ; $k \leftarrow 0$ ; $k_{c} \leftarrow k$\label{emptyQueue}\;
	  $b \leftarrow \mathcal{U}[0,2^k\rm{CW}_{min}-1]$\;
	  \While{there is a packet to transmit}{
	    \Repeat{($r = R$) or (success)}{
	      \While{$B>0$}{
	        wait 1 slot\;
	        $B \leftarrow B-1$\;
	      }
	      \colorbox{yellow}{Attempt transmission of $2^k$ packets;}\\
	      \If{collision}{
	        $r \leftarrow r+1$\;
	        $k \leftarrow \min (k+1,m)$\;
	        $B \leftarrow \mathcal{U}[0, 2^k  \rm{CW}_{min} -1]$\;
	      }
	    }
	    $r \leftarrow 0$\;
	    \eIf{success}{
	      \colorbox{yellow}{$B_{d} \leftarrow \lceil 2^{k}\rm{CW}_{min}/2\rceil-1$;}\label{backoffExample}\\
		$B \leftarrow B_{d}$\;
	    }
	    {
	      Discard $2^{k_{c}}$ packets\label{discard}\;
	      $B \leftarrow \mathcal{U}[0, 2^k \rm{CW}_{min}-1]$\;
	    }
	
	$k_{c} \leftarrow k$\;
	  }
	  Wait until there is a packet to transmit\;
	}	
	\vspace{0.2cm}
	\caption{CSMA/ECA$_{\text{Hys+FS}}$: $k_{c}$ refers to the contention backoff stage, that is, the backoff stage with which a contention for transmission is started. After $R$ retransmission attempts, Fair Share instructs the node to drop $2^{k_{c}}$ packets.}
	\label{alg:fullECA}
	\end{algorithm}
	
	\subsection{The effects of Aggregation}\label{effects-of-aggregation}
	Fair Share is an A-MPDU aggregation mechanism~\cite{A-MPDU} that coupled with the collision-free schedule built by CSMA/ECA$_{\text{Hys}}$ is able to provide short-term fairness. However, it also improves the throughput since the aggregation process makes the packet transmission more efficient by reducing overheads. Furthermore, the level of aggregation provided by Fair Share depends only on the buffer occupancy and the current backoff stage.
	
	The downside of Fair Share is that it may increase the time between two consecutive transmissions from the same node, which may affect negatively delay-sensitive applications such as gaming or high definition real-time video. An example of the duration of a transmission, $T(l_{k})$, is provided by (\ref{eq:Ts}) in the following Section~\ref{ECA-bounds}.
	
	In scenarios where short-term fairness and the time between consecutive transmissions are not relevant, Fair Share can be replaced by \emph{Maximum Aggregation} (MaxAg), which will significantly improve the system throughput. In Maximum Aggregation all nodes aggregate as many packets as possible at every transmission, i.e., they send $2^m$ packets in each attempt.

\begin{figure}[tb]
	\centering
		\includegraphics[width=0.7\linewidth,angle=-90]{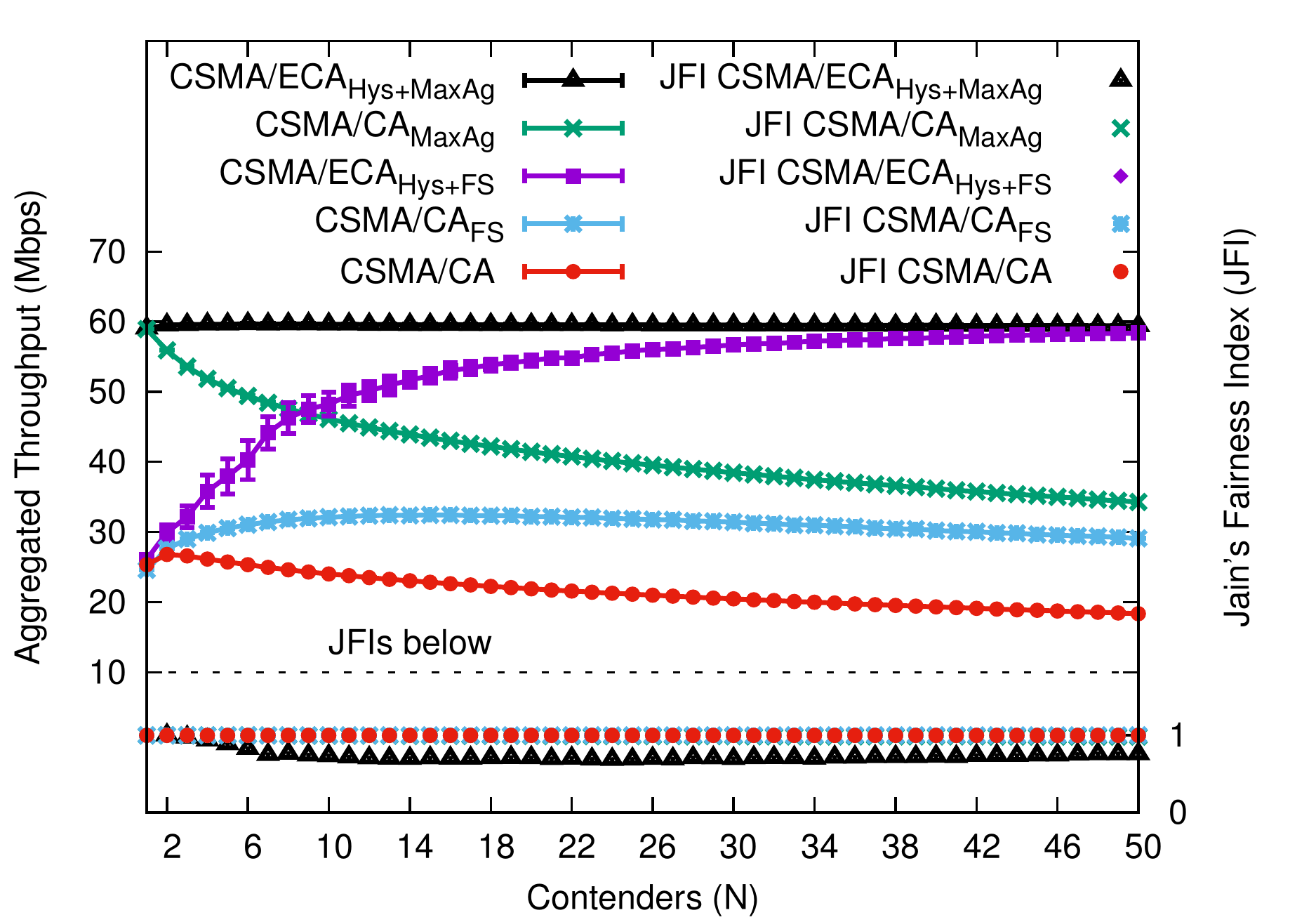}
		\caption{Throughput comparison with CSMA/CA$_{\text{MaxAg}}$: even though at low number of contenders CSMA/CA$_{\text{MaxAg}}$ achieves greater throughput than CSMA/ECA$_{\text{Hys+FS}}$, collisions eventually degrade the throughput below CSMA/ECA$_{\text{Hys+FS}}$'s when the number of contenders increases past $N=10$}
		\label{fig:ECA-vs-DCF-maxAgg}
	\end{figure}

	Figure~\ref{fig:ECA-vs-DCF-maxAgg} shows the aggregated throughput for CSMA/ECA$_{\text{Hys+FS}}$, CSMA/ECA$_{\text{Hys+MaxAg}}$, CSMA/CA using the Fair Share mechanism (CSMA/CA$_{\text{FS}}$), and CSMA/CA$_{\text{MaxAg}}$. JFIs are shown below. Although CSMA/CA$_{\text{FS}}$ and CSMA/CA$_{\text{MaxAg}}$ perform aggregation, collisions degrade the aggregated throughput as more contenders attempt transmission. On the other hand, CSMA/ECA$_{\text{Hys+FS}}$ is able to build a collision-free schedule and takes advantage of the aggregation provided by Fair Share, which opposed to just using CSMA/ECA$_{\text{Hys+MaxAg}}$, it is fair.
	
	
	To summarise the effects of using aggregation:
	\begin{itemize}
		\item It increases the aggregated throughput: because nodes are able to send multiple packet in each attempt, the system throughput is increased. Moreover, Fair Share compensates those nodes at higher backoff stages to ensure throughput fairness.
		\item Maximum aggregation supposes the deactivation of the Fair Share mechanism: performing maximum aggregation upon each transmission attempt is equivalent to having different schedule lengths and not compensating nodes at higher backoff stages. Although the aggregated throughput increases, this results in an uneven distribution of the channel time among contenders, which renders it unfair.
		\item Longer periods between transmission attempts: given that each transmission takes longer, the time between transmission attempts also increases. This may specially affect delay-sensitive applications.
	\end{itemize}

	
	Since we consider that fairness and a short inter-transmission time are even more important than raw throughput for the next generation of WLANs, we keep CSMA/ECA$_{\text{Hys+FS}}$ as the reference protocol.
	
	\subsection{Throughput bounds of CSMA/ECA$_{\text{Hys+FS}}$}\label{ECA-bounds}
	
	They correspond to the maximum and minimum achievable throughput without the possibility of collisions using Hysteresis and Fair Share. The ideal CSMA/ECA$_{\text{Hys+FS}}$ network uses the minimum schedule length that guarantees a collision-free operation. That is, with a schedule length of $C=2^{k}B_{\text{d}}+1$, where $k = \left\lceil log_{2}(N/(B_{\text{d}}+1))\right\rceil$. Using this minimum schedule length, $N$ nodes will be at the same backoff stage if $N\leq B_{\text{d}}+1$. Otherwise, $h = N-(C-N)$ nodes would occupy the $k$-th backoff stage and the other $N-h$ nodes the $(k-1)$-th one. The system throughput is computed as follows:
	
	\begin{equation}
			S =
				\begin{cases}\label{eq:bound}
					hs_{k}(l_{k}) + (N-h)s_{k-1}(l_{k-1}), & \text{if } N > B_{\text{d}}+1 \\
					Ns_{k}(l_{k}), & \text{otherwise}
				\end{cases}
	\end{equation}
	
	where $s_{k}(l_{k})$ and $s_{k-1}(l_{k-1})$ are the throughput achieved by the nodes at the $k$-th and $(k-1)$-th backoff stages sending $l_{k}$ and $l_{k-1}$ packets respectively. These are given by:
	
	\begin{subequations}
		\begin{align}
			&s_{k}(l_{k}) = \frac{l_{k} L}{hT(l_{k})+2(N-h)T(l_{k-1}) + \sigma_{e}(C - N)}\label{eq:HighNodes}\\
			&s_{k-1}(l_{k-1}) = \frac{l_{k-1} L}{(N-h)T(l_{k-1}) + k\frac{T(l_{k})}{2}}\label{eq:LowNodes}
		\end{align}
	\end{subequations}
	
	where $L$ is the data payload, $T(l_{k})$ and $T(l_{k-1})$ are the duration of the transmission of $l_{k}$ and $l_{k-1}$ packets, respectively; $\sigma_{e}$ is the duration of an empty slot. Additionally, $T(l_{k})$ derives from~(\ref{eq:Ts}):
	
	\begin{multline}\label{eq:Ts}
		T(l_{k})= \left( T_{\text{PHY}} + \left\lceil \frac{ \text{SF} + l_{k} (\text{MD}+\text{MH}+L) + \text{TB}}{L_{\text{DBPS}}}\right\rceil T_{\text{sym}} \right) \\ 
		+ \text{SIFS}+\left(T_{\text{PHY}} + \left\lceil\frac{\text{SF} + L_{\text{BA}} + \text{TB}}{L_{\text{DBPS}}} \right \rceil T_{\text{sym}} \right) + \text{DIFS} + \sigma_{e}
	\end{multline}
	
	where $T_{\text{PHY}}=32~\mu$s is the duration of the PHY-layer preamble and headers, $T_{\text{sym}}=4~\mu$s is the duration of an OFDM (Orthogonal Frequency Division Multiplexing) symbol. SF is the \emph{Service Field} ($16$ bits), $\text{MD}$ is the \textit{MPDU Delimiter} ($32$ bits), MH is the \emph{MAC header} ($288$ bits), TB is the number of \emph{Tail Bits} ($6$ bits), $L_{\text{BA}}$ is the \emph{Block-ACK} length ($256$ bits) and $L_{\text{DBPS}}=256$ is the number of bits in each OFDM symbol. SIFS, DIFS and $\sigma_{e}$ values can be found in Table~\ref{tab:mac-params}.

	The \emph{Lower-bound} is derived from considering the operation of an ideal CSMA/ECA$_{\text{Hys+FS}}$ network. Nodes use the minimum backoff stage possible and aggregate proportionally, thus yielding the minimum throughput achievable by a CSMA/ECA$_{\text{Hys+FS}}$ network. It is computed following~(\ref{eq:bound}) with $l_{k}=2^{k}$ and $l_{k-1}=2^{k-1}$.
	
	The \emph{Upper-bound} is obtained from considering the operation of a network using CSMA/ECA$_{\text{Hys+FS}}$, but forcing nodes to use the maximum backoff stage for determining the cycle length and the level of aggregation. It is also computed using~(\ref{eq:bound}) but considering that all nodes are in the maximum backoff stage ($k=m$) and therefore $l_{k}=2^{m}$.
	
\begin{figure}[tb]
	\centering
		\includegraphics[width=0.7\linewidth,angle=-90]{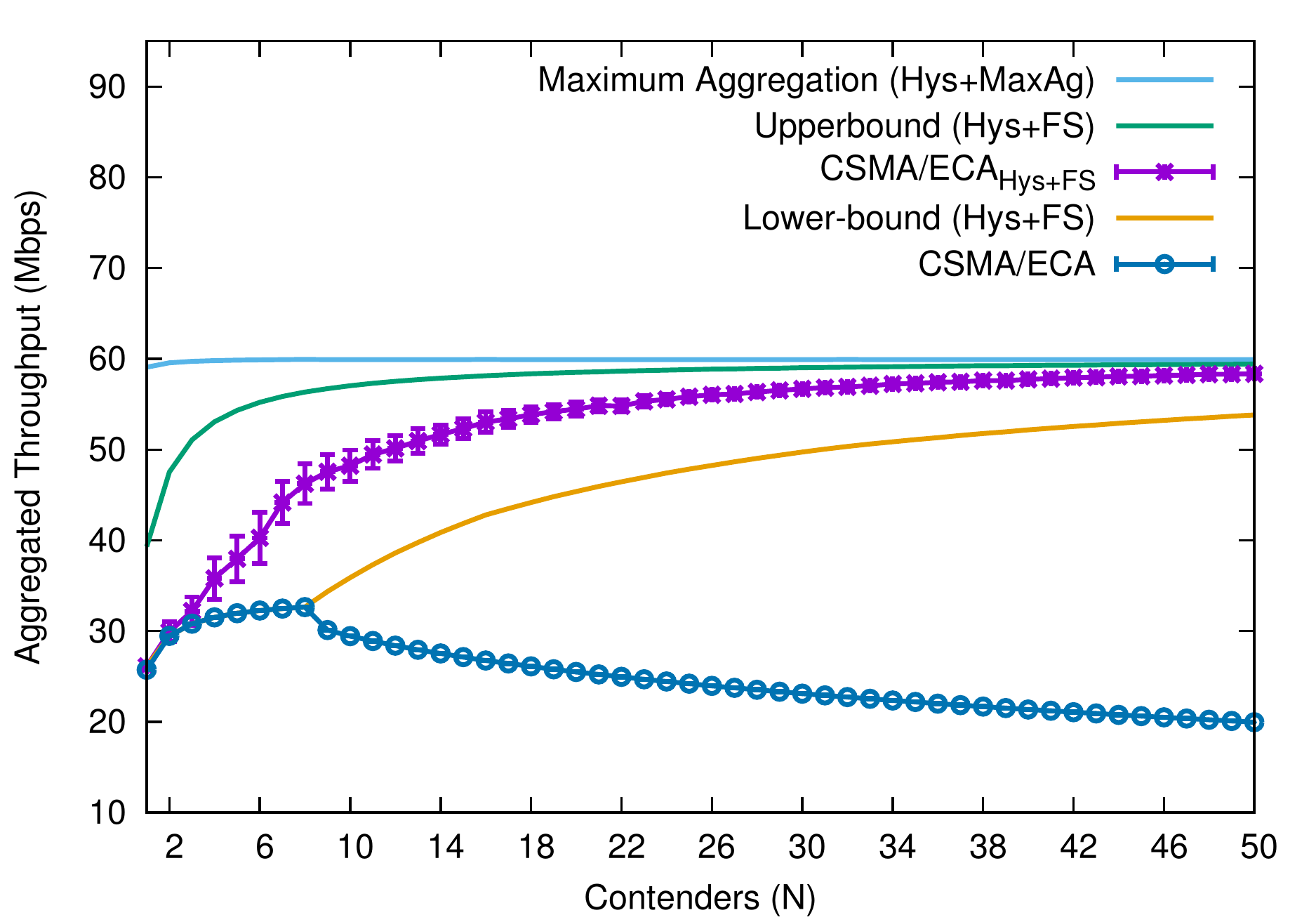}
		\caption{Upper and Lower throughput bounds for CSMA/ECA$_{\text{Hys+FS}}$}
		\label{fig:ECA-bounds-comparison}
	\end{figure}

	The maximum throughput achievable is the result of deactivating the Fair Share rules by forcing nodes to use maximum aggregation regardless of their backoff stage. This is called \emph{Maximum Aggregation (Hys+MaxAg)} in Figure~\ref{fig:ECA-bounds-comparison}. It can be derived from~(\ref{eq:bound}) considering $l_{k} = 2^{m}$ and $l_{k-1}=2^{m}$.
		
	It is interesting to see in Figure~\ref{fig:ECA-bounds-comparison} how collisions force colliding CSMA/ECA$_{\text{Hys+FS}}$ contenders to increase their backoff stage and aggregate more with Fair Share. This explains why the CSMA/ECA$_{\text{Hys+FS}}$ curve separates itself from the \emph{Lower-bound} at a very low number of contenders. 
	
	Although using maximum aggregation (see \emph{Maximum Aggregation (Hys+MaxAg)} curve in Figure~\ref{fig:ECA-bounds-comparison}) increases the throughput it carries the effect of unevenly distributing the available bandwidth among contenders, as mentioned in Section~\ref{effects-of-aggregation}.
	
	The tools for deriving these two curves are available as MATLAB functions in~\cite{ECA-bounds-example}. 
	
	\subsection{Clock drift issue in descentralized collision-free MAC protocols}\label{clockDrift-issue}
	CSMA/ECA relies on stations being able to correctly count empty slots and consequently attempt transmissions in the appropriate slot according to the backoff timer. Failure to do so may be caused by clock imperfections inside the Wireless Network Interface Cards (WNIC), which is commonly referred to as \emph{clock drift}. As pointed out in~\cite{slotDrift}, clock drift is a common issue that degrades the throughput in distributed collision-free MAC protocols like the ones reviewed in Section~\ref{relatedWork}.
	
	While miscounting empty slots have no significant effect on CSMA/CA's throughput~\cite{slotDrift}, it has a direct impact on CSMA/ECA. In a collision-free schedule with saturated CSMA/ECA contenders, a station miscounting empty slots will \emph{drift} to a possibly busy slot, collide and force a re-convergence (if possible) to a collision-free schedule (see Section~\ref{performanceClockDrift}).
	
%

	\subsection{Channel errors}\label{errorEffect}
	Failed transmissions due to channel errors are handled as collisions by CSMA/ECA$_{\text{Hys+FS}}$. Therefore, collision-free periods under this type of channel model are frequently interrupted. In order to accelerate the convergence to collision-free schedules in presence of channel errors CSMA/ECA$_{\text{Hys+FS}}$ instruct nodes to \emph{stick} to the current deterministic backoff even after \emph{stickiness} number of consecutive failed transmissions. Stickiness has been introduced to CSMA/ECA in~\cite{barcelo2011tcf}, where it is described and evaluated. Stickiness allows for a faster convergence towards a collision-free schedule, especially when performing under heavy channel errors.
	
	Failed transmissions due to channel errors also means that a few moments of operation under a noisy channel can increase the contenders's deterministic backoff to its maximum value. Such event carries the undesired effects of increasing the time between successful transmissions and reducing the overall system throughput due to fewer collision-free periods of operation.
		
	\subsection{Hidden terminals}\label{hidden}
	One key characteristic of IEEE 802.11 devices is that their carrier sense range is at least two times greater than their data range~\cite{tuningCarrierSense}. In this situation, the impact of hidden nodes can be considered to be low. This is because a given transmission could only be interfered by other transmissions from very distant nodes with energy levels not higher than the noise floor, which using modern radios can be easily solved by the capture effect. However, in specific deployments, where obstacles play also an important role on the propagation effects, hidden nodes may appear, causing asynchronous packet collisions~\cite{throughputUnderHiddenTerm}.
	
	In the case of CSMA/ECA$_{\text{Hys+FS}}$ WLANs, the same negative effects as in IEEE 802.11 WLANs are expected. In addition, collisions with hidden terminals will cause nodes to leave any collision-free schedule, as channel errors do, thus continuously disrupting any attempt of collision-free operation.
	
%
		
	\subsection{Schedule Reset Mechanism}\label{schedReset}
	CSMA/ECA$_{\text{Hys+FS}}$ instructs nodes to keep their current backoff stage after a successful transmission (resetting it to zero only if the node empties its MAC queue, see line~\ref{emptyQueue} in Algorithm~\ref{alg:fullECA}). This is done in order to increase the cycle length and provide a collision-free schedule for many contenders, which is desirable in dense scenarios.
	
	Nevertheless, having a big deterministic backoff increases the time between successful transmissions. If not operating in a scenario with many nodes, the empty slots between transmissions are not longer negligible and degrade the overall throughput of the system. For instance, if nodes withdraw from the contention (i.e.: empty their MAC queue, or move to other network) their previously used slots will now be empty. On the other hand, in scenarios with channel errors contenders rapidly end up with the largest deterministic backoff, remaining there until the MAC queue empties. Nodes should be aware of this issue and pursue opportunities to reduce their deterministic backoff, $B_{\text{d}}$ without sacrificing too much in collisions. 
	
	The \emph{Schedule Reset} mechanism for CSMA/ECA$_{\text{Hys+FS}}$ consists on finding the smallest collision-free schedule (if any) between a contender's transmissions and then change the node's deterministic backoff to fit in that schedule. Take a contender with $B_{\text{d}}=31$ as an example. By listening to the slots between its transmissions, it is possible to determine the availability of smaller (and possibly) collision-free schedules.
	
\begin{figure}[tb]
		\centering
			\includegraphics[width=\linewidth]{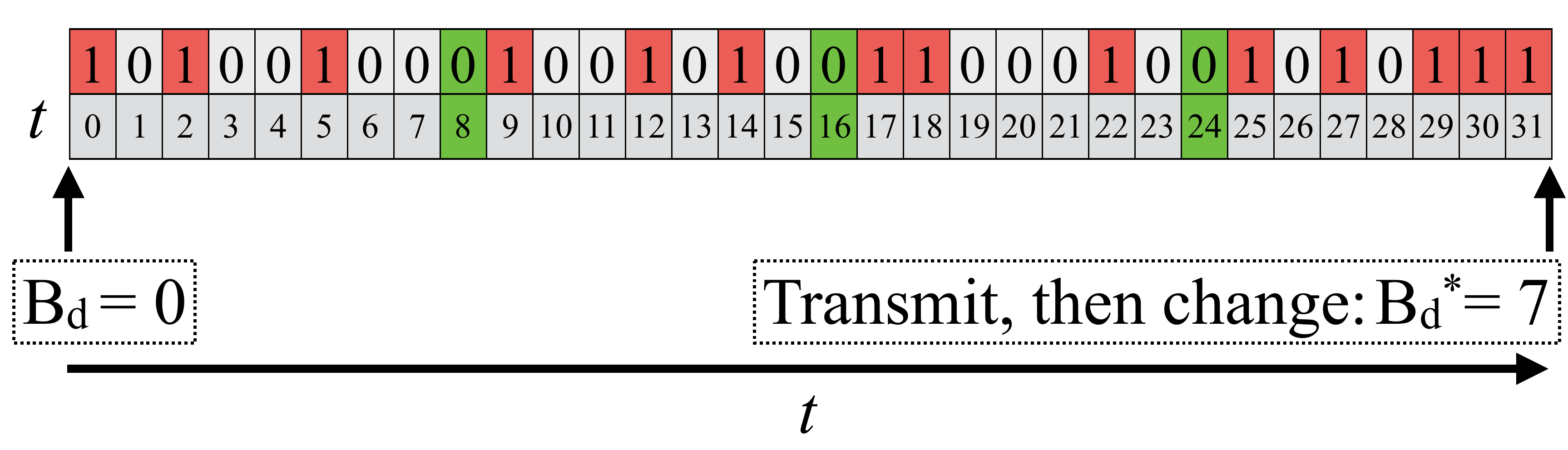}
			\caption{Example of the Schedule Reset mechanism. The lower row shows values of $t$ for a bitmap $b[t]$ of size $|b[t]| = B_{\text{d}} + 1=32$}
			\label{fig:scheduleReset1}
		\end{figure}

	Figure~\ref{fig:scheduleReset1} shows the slots between the transmissions of a contender with $B_{\text{d}}=31$. Starting from the left, the current $B_{\text{d}}=0$ means that the first slot will be filled with the contender's own transmission. Each following slot containing either a transmission or a collision is identified with the number one, while empty slots are marked with a zero. Notice that the highlighted empty slots appear every eight slots, suggesting that a schedule reduction from $B_{\text{d}}=31$ to $B_{\text{d}}^{*}=7$ is possible; where $B_{\text{d}}^{*}$ is the new deterministic backoff assigned by Schedule Reset\footnote{$B_{\text{d}}=2^{0}CW_{\min}/2-1=7$ (see Table~\ref{tab:mac-params}). So the change is made to the backoff stage, $k$. In this case from $2$, to $k=0$. This means that any new schedule must also be a power of two.}. The schedule change is performed after the contender's next successful transmission.
	
			\begin{algorithm}[tb]
		\If{sxTx == 1}{
			Initialising\;
			\If{B == 0}{
				$b[B_{\text{d}}+1] = \{0\}$\;
				$t = 0$\;
			}
		}
		\If{(sxTx $>$ 0) \&\& (sxTx $\leq~\gamma$)}{
			Filling the bitmap\;
			$b[t] \|\sigma_{i}(t)$\label{bitwiseOR}\;
			$t++$\;
			\If{$t == B_{\text{d}}$}{
				$t=0$\;
			}
		}
		\If{sxTx == $\gamma$}{\label{evaluations}
			Analising\;
			sxTx = $0$\;
			\For{($j=0;~j<k;~j++$)}{\label{sets}
				$y=\lceil 2^{j}CW_{\min}/2\rceil$\;
				isItPossible $= 0$\;
				\For{($m=y;~m<B_{\text{d}}+1;~m+=y$)}{
					isItPossible$~+=b[m]$\;\label{filling}
				}
				\If{isItPossible == $0$}{\label{testFilling}
					change = $1$\;
					newStage = $m$\;
					{\bf break};
				}
			}
		}
		\If{change == $1$}{
			Making the change\;
			$k$ = newStage\;
			change = $0$\;\label{endAssignment}
		}
		\vspace{0.2cm}
		\caption{Schedule Reset Mechanism for CSMA/ECA$_{\text{Hys+FS}}$. Every consecutive successful transmission increases the variable SxTx by one, while a collision resets it to zero. The algorithm is called a \emph{reset} when all smaller deterministic backoff are tested. On the other hand, when $j$ in line~\ref{sets} is initialised to $j=k-1$, it is called a schedule \emph{halving}}\label{alg:schedRest}
	\end{algorithm}

		\subsubsection{A conservative schedule reduction}
		Schedule Reset is described in Algorithm~\ref{alg:schedRest}. Each contender starts filling a bitmap $b$ of size $B_{\text{d}}+1$ with the information observed in each passing slot during $\gamma=\lceil C/B_{\text{d}}\rceil$ consecutive successful transmissions (sxTx in Algorithm~\ref{alg:schedRest}); where $C=\lceil 2^{m}CW_{\min}/2\rceil-1$ is the biggest deterministic backoff, and $\gamma$ is referred to as the Schedule Reset \emph{threshold} from this point forward.
		
		Each bit $t,~t\in\{0,\ldots ,B_{\text{d}}\}$ in the bitmap is the result of a bitwise OR operation between its current value and the state of the corresponding slot. This is shown in Line~\ref{bitwiseOR} in Algorithm~\ref{alg:schedRest}, where $\sigma_{i}(t)$ is a function that evaluates to $0$ if the slot corresponding to $b[t]$ is empty in iteration $i$, $i \in\{1,\ldots,\gamma\}$; or $1$ otherwise. After $\gamma$ iterations, the bitmap is evaluated (line~\ref{evaluations} in Algorithm~\ref{alg:schedRest}).
		
		
		
		Schedule Reset tests all possible deterministic backoffs that are smaller than the current $B_{\text{d}}$ (lines~\ref{sets}-\ref{filling} in Algorithm~\ref{alg:schedRest}), starting with the smallest one. If the corresponding bits in the bitmap are registered as empty, the process is stopped and the change to the new deterministic backoff is made (lines~\ref{testFilling}-\ref{endAssignment} in Algorithm~\ref{alg:schedRest}). Otherwise, the process is restarted after the next successful transmission.
		
		Using Figure~\ref{fig:scheduleReset1} as an example, we can see that $\sum\limits_{t} b[t]=0$, for $t\in\{8,16,24\}$. This means that the transmission slots corresponding to a deterministic backoff, $B^{*}_{\text{d}}=7$ are free, therefore a change of schedule is possible. In case of suffering a collision immediately after applying the schedule reduction, the node reverses the changes made by Schedule Reset before handling the collision.
		
	
		\subsubsection{Aggressive approach to face channel errors}\label{aggr}
		
		the default $\gamma$ ensures that the bitmap registers all the transmission slots in the network (assuming saturated traffic and a perfect channel), providing enough information for performing the schedule reduction without disrupting collision-free schedules. 
		
		Nevertheless, this $\gamma$ can be rendered too conservative and unnecessary because nodes randomly leave the collision-free schedule due to errors. This condition produces Schedule Reset bitmaps that do not contain enough information for successfully avoiding collisions after the schedule reduction. This is also true for any $\gamma^{*}>\gamma$.
		
		This effect is leveraged by setting $\gamma=1$, meaning that the bitmap is filled with the information of slots between only two consecutive transmissions. This measure increases the frequency of schedule reset attempts. Furthermore, incrementing the node's stickiness after a successful reduction of the schedule allows for a faster convergence towards a collision-free schedule, especially when performing under heavy channel errors. 
	
	\subsection{Backwards compatibilty and coexistence}
	CSMA/ECA$_{\text{Hys+FS}}$ springs from a modification to CSMA/CA's backoff mechanism. It keeps the range of values CSMA/CA nodes use to draw a random backoff (i.e., use the same $CW_{\min}$ and $CW_{\max}$), allowing CSMA/ECA$_{\text{Hys+FS}}$ contenders to coexist with CSMA/CA nodes in the same network. Further, the selection of CSMA/ECA$_{\text{Hys+FS}}$'s deterministic backoff, $B_{\text{d}}$, is the expected value for the current backoff stage $k$ ($B_{\text{d}}\coloneqq\lceil{E[0,CW(k)-1]}\rceil$)~\cite{research2standards}, which ensures fairness among contenders. An overview of the attained throughput for different proportions of CSMA/ECA$_{\text{Hys+FS}}$ and CSMA/CA nodes is presented in Section~\ref{coexistance-w-csmaca}.\section{Simulation Scenario}\label{simulations}
This section provides the simulation parameters for testing CSMA/ECA$_{\text{Hys+FS}}$ under two different traffic conditions, namely saturated and non-saturated. We also provide details on how channel errors are modelled and what are its effects over the transmissions. Further, the simulation of the clock drift effect, Schedule Reset parameters, and the coexistence with CSMA/CA are also subjects to be addressed in this section.

	\subsection{Scenario details}
	Results are obtained by running multiple simulations over a modified version of the COST simulator~\cite{COST}, available at~\cite{sim:parameters-TON}. PHY and MAC parameters are detailed in Table~\ref{tab:mac-params}. Some assumptions were made in order to test the performance at the MAC layer:
	
	\begin{itemize}
		\item Unspecified parameters follow the IEEE 802.11n ($2.4$~GHz) amendment.
		\item All nodes are in communication range.
		\item No Request-to-Send (RTS) or Clear-to-Send (CTS) messages are used.
		\item Collisions take as much channel time as successful transmissions.
	\end{itemize}
	
	The aforementioned assumptions ensure that the simulation results are just effects of the MAC behaviour.
	If not mentioned otherwise, results are derived from 20 simulations of 100 seconds in length, each one with a different seed. Figures also show the standard deviation.
	
	\begin{table}
		\centering
		\caption{PHY and MAC parameters for the simulations}
		\label{tab:mac-params}
		\begin{tabular}{|c|c|}
			\hline
			\multicolumn{2}{|c|}{{\bfseries PHY}}\\
			\hline
			{\bfseries Parameter} & {\bfseries Value}\\
			\hline
			PHY rate & 65~Mbps\\
			Empty slot & $9~\mu s$\\
			DIFS & $28~\mu s$\\
			SIFS & $10~\mu s$\\
			\hline
			\multicolumn{2}{|c|}{{\bfseries MAC}}\\
			\hline
			{\bfseries Parameter} & {\bfseries Value}\\
			\hline
			Maximum backoff stage ($m$) & 5\\
			Minium Contention Window ($CW_{\min}$) & 16\\
			Maximum retransmission attempts & 6\\
			Data payload (Bytes) & 1024\\
			MAC queue size (Packets) & 1000\\
			\hline
		\end{tabular}
	\end{table}
	
	\subsection{Saturated and Non-saturated stations}\label{unsaturation}
	A saturated station always has packets in its MAC queue. This is modelled by setting the packet arrival rate to the MAC queue ($\Delta_{\text{PAR}}$) to a value greater than the achievable throughput. To ensure saturation, stations are set to fill their MAC queue at $\Delta_{\text{PAR}}=65$~Mbps, which is purposefully greater than the effective capacity of the channel.
	
	To evaluate the performance under non-saturated conditions, stations need to be able to empty their MAC queues. To do so, the packet arrival rate to the MAC queue is set to $\Delta_{\text{PAR}}=1$~Mbps. These values of $\Delta_{\text{PAR}}$ have proven to produce the desired effects.
	
	\subsection{Performance under clock drift}
	Clock drift is simulated by setting a drift probability, $p_{cd}$. Each station has a probability of $p_{cd}/2$ of miscounting one slot more, and $p_{cd}/2$ of miscounting one slot less. This approach follows the one proposed by Gong et. al in~\cite{slotDrift}.
	
	\subsection{Channel errors}\label{channelErrorsDef}
	Channel errors are modeled by assigning a probability of a packet being corrupted by the channel, $p_e>0$. That is, in a single packet transmission there is probability $p_e$ that the transmission will not be acknowledged. If the transmission is an A-MPDU (like in the case of CSMA/ECA$_{\text{Hys+FS}}$), $p_e$ will affect each MPDU individually and independently. Therefore, an A-MPDU transmission will be considered a complete failure only if all frames in the aggregation are affected by the channel error probability.
	
	All results are shown with stickiness equal to one (see Section~\ref{errorEffect}). That is, after a colision, CSMA/ECA$_{\text{Hys+FS}}$ nodes will use a random backofff. Nevertheless, in Section~\ref{resultsSchedRest} curves described as \emph{dynStick} temporarily increase the node's stickiness to two after a successful reduction of the schedule (using a random backoff after two consecutive collisions). This increase is done in order to converge faster to collision-free schedules when operating with channel errors.
	
	\subsection{Coexistance with CSMA/CA}\label{coexistence}
	To test the performance of CSMA/CA and CSMA/ECA$_{\text{Hys+FS}}$ stations in the same network, simulations are set with a CSMA/CA node density of 1/4, 1/2 and 3/4 of the total.
	
	\subsection{Applying Schedule Reset} 
	A set of results under saturated conditions and channel errors applying Schedule Reset (see Section~\ref{schedReset}) are provided. Some of the results are generated with a $\gamma=1$, or \emph{aggressive} Schedule Reset (indicated as \emph{aggr.} in the figures). These settings provide the highest throughput under the tested conditions, and also in the real hardware implementations such as the one shown in Section~\ref{EDCA}.

\section{Results}\label{results}
	\begin{figure*}[tb]
		\centering
		\includegraphics[width=0.5\linewidth,angle=-90]{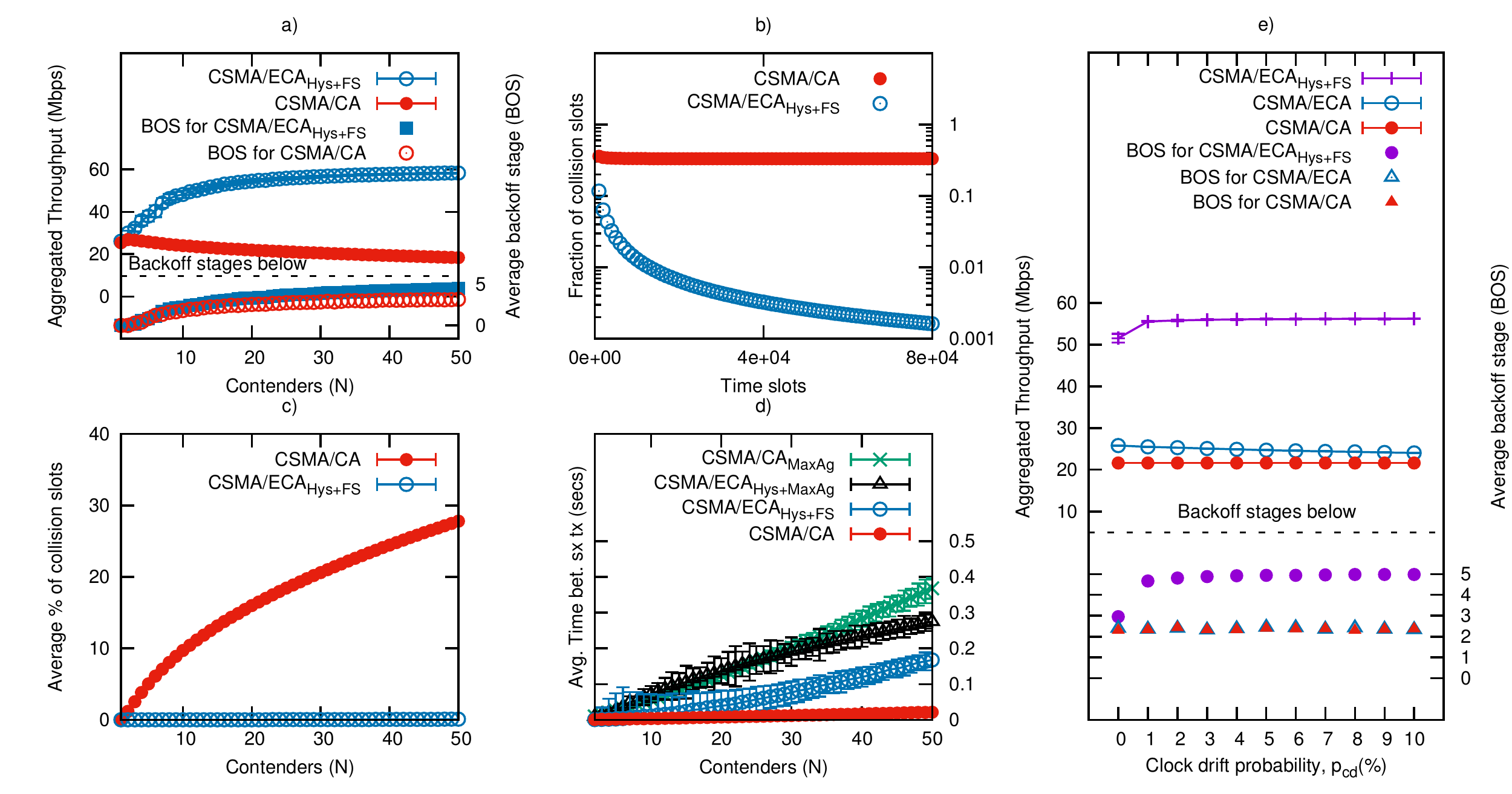}
		\caption{Simulation results under saturated traffic: a) Throughput under saturated conditions; b) Evolution of the fraction of collision slots in a scenario with 70 saturated stations; c) Average percentage of collision slots: fraction of time slots containing collisions; d) Average time between successful transmissions (sx tx), e) Throughput when increasing the clock drift probability}
		\label{fig:satResults}
	\end{figure*}
	
	

	\subsection{Saturated nodes}\label{resultsSaturated}
	In CSMA/CA, a large number of saturated nodes will normally be related to a high collision probability. This effect is in part the result of resetting the backoff stage after a successful transmission and the generation of a new random backoff. However, this scenario provides an advantageous condition to CSMA/ECA$_{\text{Hys+FS}}$ nodes. In saturation, CSMA/ECA$_{\text{Hys+FS}}$ nodes build a collision-free schedule and stick to their deterministic backoff as long as they transmit successfully, effectively eliminating collisions.
	
	This section aims at overviewing the throughput of CSMA/CA and CSMA/ECA$_{\text{Hys+FS}}$ in saturation, as well as the collision probability, the average time between successful transmissions and the effect of clock drift over the throughput.
	\\
	\subsubsection{Throughput}
	CSMA/ECA$_{\text{Hys+FS}}$ nodes are able to build a collision-free schedule, use the channel more efficiently, and experience a throughput increase as seen in Figure~\ref{fig:satResults}a. Hysteresis allows the allocation of more contenders by increasing the length of the collision-free schedule, while Fair Share ensures an even distribution of the available throughput. This is reflected by the average backoff stage, which value increases with the number of contenders. In contrast, CSMA/CA throughput keeps decreasing due to an augmented number of collisions as the number of nodes increases (see Figure~\ref{fig:satResults}c). Further, Figure~\ref{fig:satResults}b shows the fraction of collision slots for CSMA/ECA$_{\text{Hys+FS}}$ and CSMA/CA as simulation time passes. In the figure, is appreciated how the fraction of collision slots keeps decreasing once CSMA/ECA$_{\text{Hys+FS}}$ reaches collision-free operation.

	\subsubsection{Effect of clock drift over the achieved throughput in saturation}\label{performanceClockDrift}
	Figure~\ref{fig:satResults}e shows the network aggregated throughput with 16 saturated stations and an increasing clock drift probability.
	
	
	In Figure~\ref{fig:satResults}e, a station has a clock drift probability equal to $p_{cd}$. Each station has a probability of $p_{cd}/2$ of miscounting one slot more, and $p_{cd}/2$ of miscounting one slot less. Because CSMA/CA is based on a random backoff, miscounting slots has no significant effect on the throughput. For the CSMA/ECA curve, it is possible to appreciate a slight decrease of the throughput as $p_{cd}$ increases, caused by collisions due to the drift.
	
	The CSMA/ECA$_{\text{Hys+FS}}$ curve in Figure~\ref{fig:satResults}e shows instead an increase of the aggregated throughput as $p_{cd}$ grows. Collisions make CSMA/ECA$_{\text{Hys+FS}}$ contenders to increment their backoff stage and aggregate packets for transmissions according to Fair Share, effectively increasing the throughput. 
	
	As it also can be appreciated in the figure, the average backoff stage for CSMA/ECA$_{\text{Hys+FS}}$ contenders increases rapidly to its maximum value ($m=5$), reducing the effect of clock drift over CSMA/ECA$_{\text{Hys+FS}}$ nodes given that their transmissions would now be separated by more slots.\\
	
	\subsubsection{Time between successful transmissions}\label{timeBetweenSxTx}
	It is related to the elapsed time between the contention for transmission and the reception of an ACK.

	
	In Figure~\ref{fig:satResults}d all tests with maximum aggregation, namely CSMA/CA$_{\text{MaxAg}}$ and CSMA/ECA$_{\text{Hys+MaxAg}}$, have an increased average time between successful transmissions. This is due to the multiple packets that are sent in each attempt. CSMA/CA$_{\text{MaxAg}}$, though, has an increased value due to collisions also taking longer channel time.
	
	Although CSMA/ECA$_{\text{Hys+FS}}$ has an increased average time between successful transmissions due to Fair Share, it has a lower metric when compared with the maximum aggregation curves in Figure~\ref{fig:satResults}d.


	\subsection{Non-saturated nodes}\label{resultsUnsaturated}
	

	Emptying the MAC queue in CSMA/ECA$_{\text{Hys+FS}}$ means that nodes will reset their backoff stage to zero and use a random backoff when a new packet arrives at the queue, breaking any collision-free operation in CSMA/ECA$_{\text{Hys+FS}}$. The following show the impact over throughput, delay and time between successful transmissions when using CSMA/CA and CSMA/ECA$_{\text{Hys+FS}}$ in non-saturated conditions.\\
	
	\subsubsection{Throughput}
	
\begin{figure*}[tb]
		\centering
		\includegraphics[width=0.5\linewidth,angle=-90]{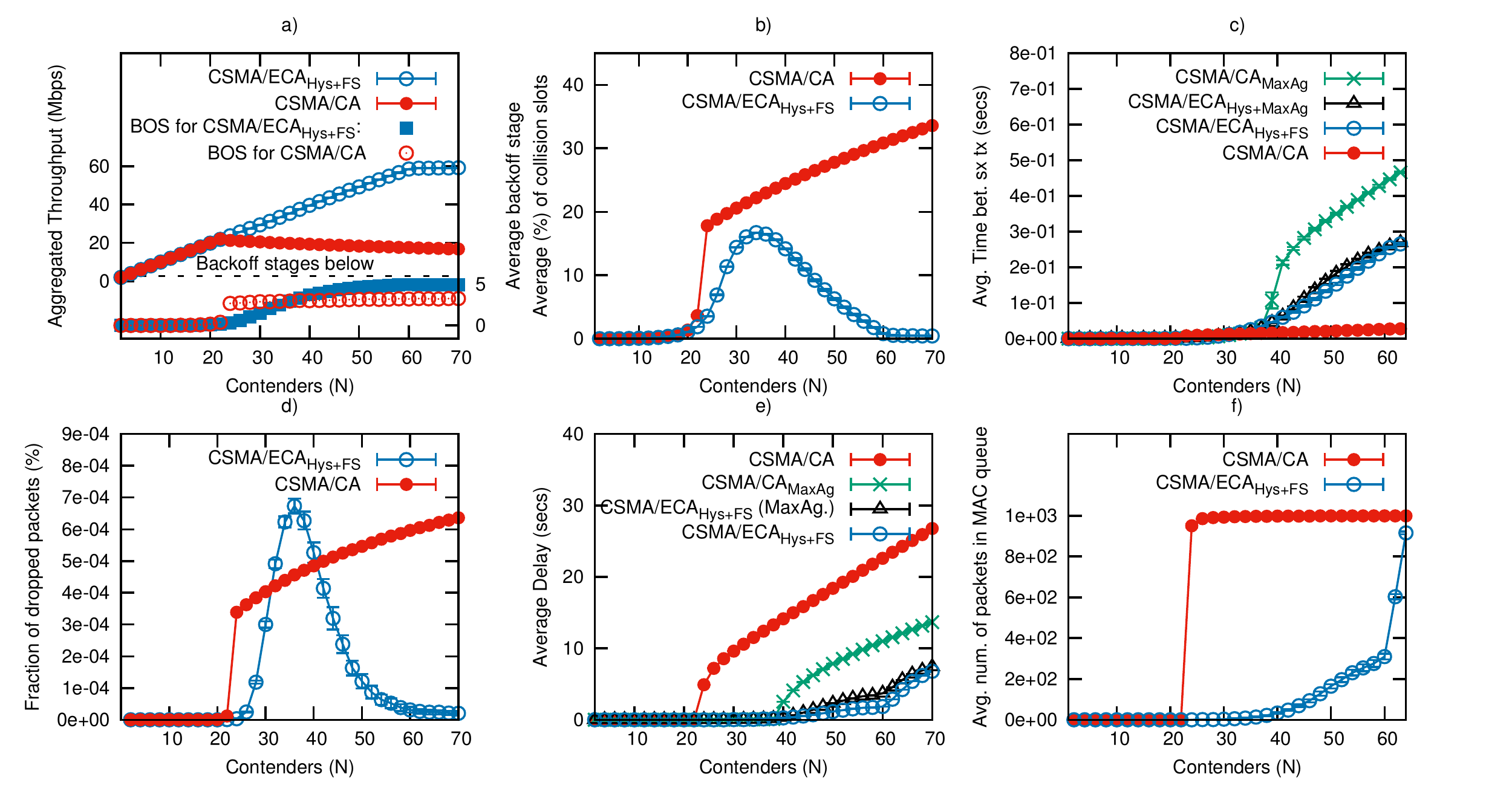}
		\caption{Simulation results under non-saturated traffic: a) Throughput; b) Average percentage of collision slots: the fraction of time slots containing collisions; c) Average time between successful transmissions; d) Average fraction of dropped packets; e) Average system delay; f) Average number of packets in the MAC queue of a node}
		\label{fig:unsatResults}
	\end{figure*}

	
	In Figure~\ref{fig:unsatResults}a, the aggregated throughput increases linearly for the \emph{CSMA/CA} curve until saturation is reached at around 22 nodes, where the throughput begins to degrade. The \emph{CSMA/ECA$_{\text{Hys+FS}}$} curve has a similar behavior, entering saturation at around 60 nodes. Further, at around 30 nodes we see an increase in the average backoff stage for CSMA/ECA$_{\text{Hys+FS}}$ contenders which suggests an increment in collisions. This effect is shown in Figure~\ref{fig:unsatResults}b and Figure~\ref{fig:unsatResults}d, where at around 35 nodes CSMA/ECA$_{\text{Hys+FS}}$ contenders start colliding and dropping packets. 
	
	As indicated by Figure~\ref{fig:unsatResults}b, when $20<N\leq 35$ CSMA/ECA$_{\text{Hys+FS}}$ nodes suffer from an increasing number of  collisions. This is due to nodes emptying their MAC queue quicker due to Fair Share, as shown in Figure~\ref{fig:unsatResults}f and re-entering the contention with a random backoff every time a new packet arrives.
	
	This increase in collisions may also cause the dropping of packets due to reaching the maximum retransmission limit. As CSMA/ECA$_{\text{Hys+FS}}$ drops more packets after reaching such limit (see line~\ref{discard} in Algorithm~\ref{alg:fullECA}), it shows a higher fraction of dropped packets in Figure~\ref{fig:unsatResults}d.
	
	Beyond 35 contenders, the MAC queue of CSMA/ECA$_{\text{Hys+FS}}$ nodes starts to fill up, gradually allowing longer periods of collision-free operation due to CSMA/ECA$_{\text{Hys+FS}}$ nodes getting saturated. This allows CSMA/ECA$_{\text{Hys+FS}}$ to outperform CSMA/CA.\\
	
	\subsubsection{Delay}
	
	
	This metric refers to the elapsed time between the injection of a packet into the station's MAC queue and the reception of an ACK for such packet. 
	
	In Figure~\ref{fig:unsatResults}e, a rapid increase in the delay for CSMA/CA nodes is appreciated at the saturation point (around 20 contenders), whereas CSMA/ECA$_{\text{Hys+FS}}$'s delay is still low. 
	
\begin{figure}[tb]
		\centering
		\includegraphics[width=0.7\linewidth,angle=-90]{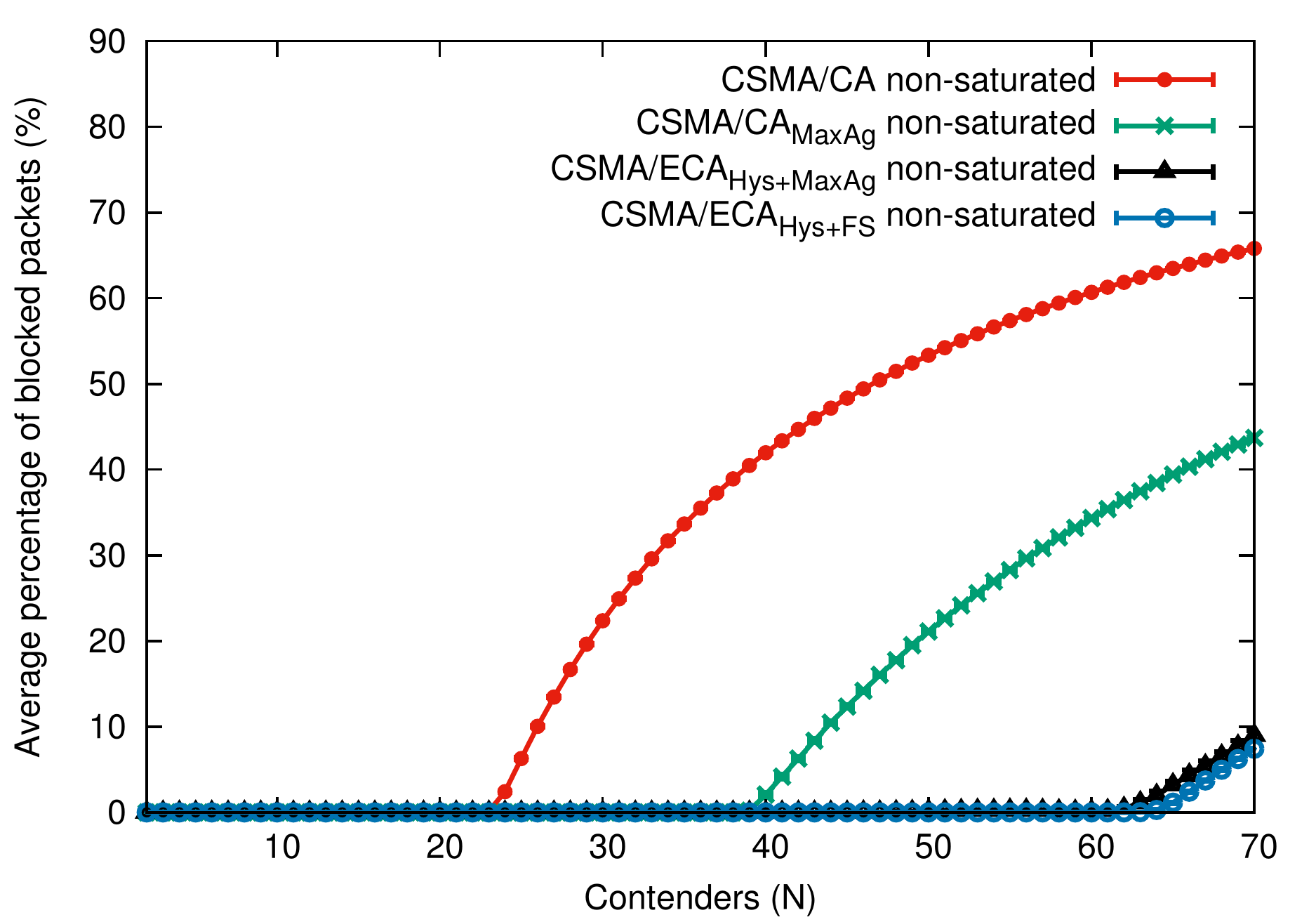}
		\caption{Average fraction of blocked packets. When the MAC queue is full, packets comming from the uppers are discarded (or blocked) at the entrance of the buffer}
		\label{fig:blocked-packets}
	\end{figure}

	Further, with CSMA/ECA$_{\text{Hys+FS}}$ the percentage of blocked packets from the MAC queue is lower than CSMA/CA or CSMA/CA$_{\text{MaxAg}}$ (see Figure~\ref{fig:blocked-packets}). This is due to the construction of collision-free schedules which ensure that large A-MPDU transmissions do not suffer from collisions.
	
	As CSMA/ECA$_{\text{Hys+FS}}$ nodes get saturated, the delay increases due to longer queueing and contention time (see the number of packets in the MAC queue for CSMA/ECA$_{\text{Hys+FS}}$ nodes in Figure~\ref{fig:unsatResults}f and how it is related to the increase in delay shown in Figure~\ref{fig:unsatResults}e).
	
	
	Figure~\ref{fig:unsatResults}c shows the average time between successful transmissions. It is possible to see from the figure that when CSMA/ECA$_{\text{Hys+FS}}$ approaches the saturation point the average time between successful transmissions increases, resembling Figure~\ref{fig:satResults}d. 
	
%
	
	\subsection{Coexistence with CSMA/CA}\label{coexistance-w-csmaca}
	
	\begin{figure*}[tb]
		\centering
		\includegraphics[width=0.5\linewidth,angle=-90]{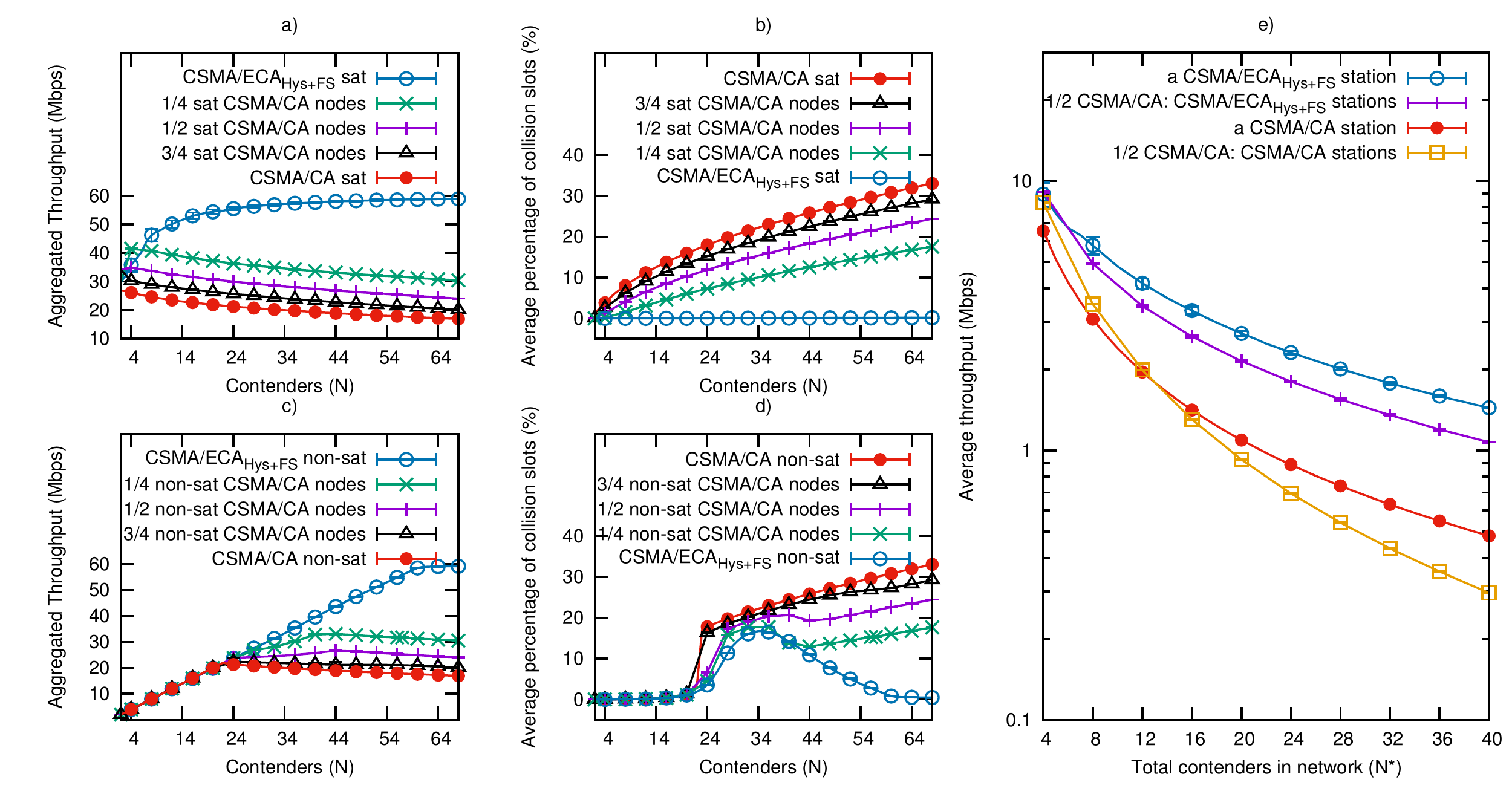}
	\caption{Coexistence results: a) Network throughput when composed by various proportions of saturated CSMA/CA and CSMA/ECA$_{\text{Hys+FS}}$ nodes; b) Average percentage of collision slots for the tested saturated mixed network setups proportions; c) Network throughput when composed by various proportions of non-saturated CSMA/CA and CSMA/ECA$_{\text{Hys+FS}}$ nodes; d) Average percentage of collision slots for the tested mixed network setups proportions in non-saturated conditions; e) Average station throughput per MAC protocol in a saturated mixed network}
		\label{fig:coexResults}
	\end{figure*}

	CSMA/ECA is thought to be an evolution of CSMA/CA given its similarities and the ability to coexists with the latter. This section provides simulations results for a setup of different proportions of CSMA/CA nodes in a network where there are also CSMA/ECA$_{\text{Hys+FS}}$ contenders, that is: 1/4, 1/2 and 3/4 of the total nodes run CSMA/CA, while the rest uses CSMA/ECA$_{\text{Hys+FS}}$. This network configuration will be referred to as \emph{mixed network setup} from here on.	
	\subsubsection{Throughput}
	
	Figure~\ref{fig:coexResults}a shows the network throughput for different proportions of CSMA/CA nodes in a mixed network setup.

	In the figure it is appreciated how the mixed network setups curves lay between the CSMA/CA and CSMA/ECA$_{\text{Hys+FS}}$ curves. As the proportion of CSMA/CA nodes decreases, the throughput increases as the result of a lower probability of collision, as can be seen in Figure~\ref{fig:coexResults}b. A similar behaviour is observed when testing the same proportion of nodes under non-saturated conditions. Figure~\ref{fig:coexResults}c and Figure~\ref{fig:coexResults}d show the average aggregated throughput and fraction of collisions slots in a non-saturated mixed network setup.
	
	As shown in Figure~\ref{fig:coexResults}a and Figure~\ref{fig:coexResults}c, at a lower proportion of CSMA/CA nodes ($1/4$) the average aggregated throughput is the greatest among the tested mixed network setups. This is because collisions trigger Hysteresis and Fair Share in CSMA/ECA$_{\text{Hys+FS}}$ nodes, lowering the number of times these nodes enter in a contention and reducing the overall collision probability when compared to an only CSMA/CA network (see Figure~\ref{fig:coexResults}b, Figure~\ref{fig:coexResults}d and Figure~\ref{fig:coexResults}e). As the proportion of CSMA/CA nodes increases, the network throughput approximates to CSMA/CA.
	
	Figure~\ref{fig:coexResults}e shows the average throughput for CSMA/CA and CSMA/ECA$_{\text{Hys+FS}}$ stations in a saturated mixed setup (half the contenders using CSMA/CA). It is possible to see that for a low total number contenders ($N^{*}\leq 12$) CSMA/CA stations attain greater throughput than in a CSMA/CA-only network. Again, this is because in the mixed network setup the other $N^{*}/2$ contenders with CSMA/ECA$_{\text{Hys+FS}}$ use a deterministic backoff, leaving many empty slots between transmissions.
	
	Still referring to Figure~\ref{fig:coexResults}e, for $N^{*}>12$ periods of collision-free operation and the aggregation performed with Fair Share allows CSMA/ECA$_{\text{Hys+FS}}$ to have larger channel time than CSMA/CA nodes, which throughput degrades even more than in CSMA/CA-only. These results suggests that a switch to CSMA/ECA$_{\text{Hys+FS}}$ can be beneficial for networks with high number of contenders.

	
	\subsection{Channel errors and Schedule Reset}\label{resultsSchedRest}
	
	\begin{figure*}[tb]
		\centering
		\includegraphics[width=0.65\linewidth,angle=-90]{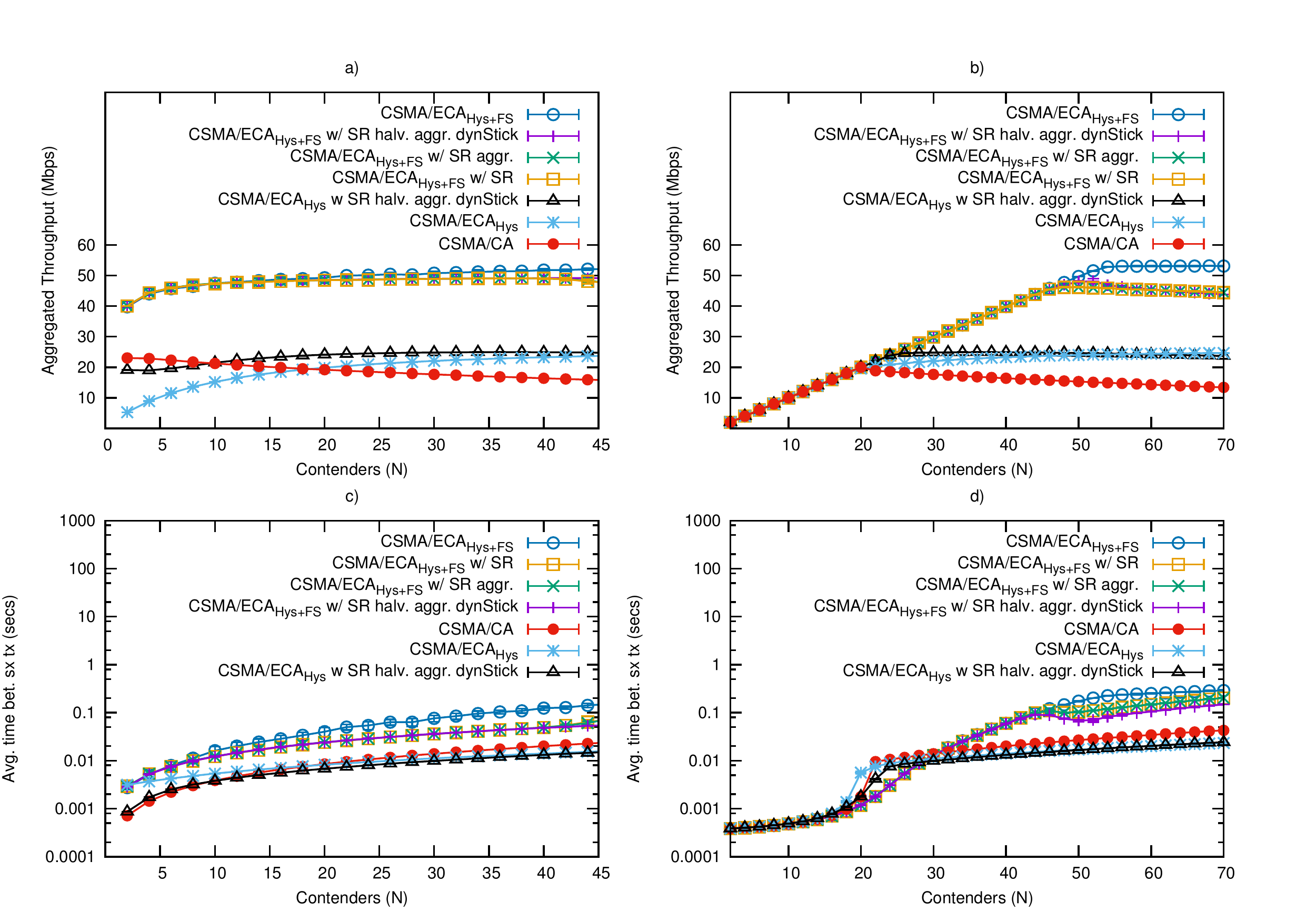}
		\caption{Performance results with Schedule Reset. Curves called "aggr" reffer to values of $\gamma=1$, while "halv" refers to a schedule halving (see Section~\ref{aggr} and Algorithm~\ref{alg:schedRest}). \emph{dynStick} reffers to curvers where the node's stickiness is temporarily increased to two after a successful reduction of the schedule: a) Average throughput for a saturated network; b) Average throughput for a non-saturated network; c) Time between successful transmissions for a saturated network; d) Time between successful transmissions for a non-saturated network}
		\label{fig:SchedResetResults}
	\end{figure*}
	
	Figure~\ref{fig:SchedResetResults}a shows the average aggregated throughput of a saturated network with a channel error probability, $p_e=0.1$ (see Section~\ref{channelErrorsDef}). We can see that the highest throughput, corresponding to CSMA/ECA$_{\text{Hys+FS}}$ is lower than the one observed in a perfect channel (Figure~\ref{fig:satResults}a). Furthermore, curves with Schedule Reset (SR) seem to have lower aggregated throughput. This is because the backoff stage, and therefore the level of aggregation done by Fair Share, is repeatedly reduced.
	
	Still assessing Figure~\ref{fig:SchedResetResults}a and focusing on the SR curves, we can see that both aggressive schedule halving with dynamic stickiness (\emph{SR aggr. halv. dynStick}) curves (with and without Fair Share) have higher aggregated throughput than the rest (for a description of aggressive schedule halving and dynamic stickiness, refer to Algorithm~\ref{alg:schedRest} and Section~\ref{aggr}). This Schedule Reset configuration reduces the time between successful transmissions (see Figure~\ref{fig:SchedResetResults}c) and maintains collision-free operation for longer periods of time thanks to the increase in the node's stickiness.
	
	Figure~\ref{fig:SchedResetResults}c shows the average time between successful transmissions for the same network setup. Even-though CSMA/ECA$_{\text{Hys+FS}}$ still has a higher metric than CSMA/CA due to the aggregation done by Fair Share, Schedule Reset is able of reducing this metric by almost $43\%$.
	
	Figure~\ref{fig:SchedResetResults}b and Figure~\ref{fig:SchedResetResults}d show the aggregated throughput and time between successful transmissions, respectively for a non-saturated network with the same $p_e=0.1$. Since under non-saturated conditions CSMA/ECA$_{\text{Hys+FS}}$ nodes reset their schedule every time they empty their MAC queues, Schedule Reset's benefits are not relevant. Nevertheless, a reduction in the time between successful transmissions is observed for Schedule Reset curves in Figure~\ref{fig:SchedResetResults}d.

%
\section{Real hardware implementation}\label{EDCA}

We confirmed the simulations results experimentally with a testbed of real prototypes. This phase was mandatory in order to check whether the behaviour of the system is still correct in the presence of real phenomena, like excessive channel errors, or technical limitations of the hardware like imperfect node synchronisation; and to verify the implementation feasibility of CSMA/ECA$_{\text{Hys}}$ and the Schedule Reset mechanism with a real time horizon.

As CSMA/ECA$_{\text{Hys}}$ works with the standard 802.11 PHY, we do not need the flexibility offered by complex development kits based on FPGAs like the WARP boards~\cite{amiri2007warp}. Any architecture that allows a customisation of the channel access delay{black} would fit our requirements. For this reason we chose a widely available 802.11b/g WNIC from Broadcom. Like many other cards, its behaviour is ruled by a firmware code that controls the underlying hardware (including the radio, carrier sense and FIFOs) in real time. In particular we used the BRCM4318KBFG PCI card as it is supported by the OpenFWWF~\cite{OpenFWWF} open-source firmware, an alternative to the original code from the manufacturer that has been recently considered as development platform in several research projects~\cite{WMP,gringolitmc14,gringoliccr14,berger14,CF-MAC}. 

Thanks to the extremely low price of this card (refurbished devices are widely available for less than $\$10$ each), it is possible to run experiments with very large testbeds. We used 25 PC-Engines Alix 2d2 nodes running Linux kernel 2.6.32 as stations and a more powerful computer with the same kernel as Access Point (AP), all nodes being equipped with the aforementioned WNIC.

We built on the DCF protocol implemented in OpenFWWF and adapted the firmware to create collision-free schedules as described in Algorithm \ref{alg:CSMA_ECA}. The modification was straightforward: for every transmitted data frame the firmware sets an ACK-time-out alarm. Later, either at the expiration of the timer or when the acknowledgment frame is received, it executes a handler labelled {\tt tx\_contention\_params\_update}. It updates the contention window value {\tt STATE\_CW} and backoff counter according to the success/failure of the previous transmission attempt, just as in Algorithm~\ref{alg:CSMA_ECA}. Implementing CSMA/ECA$_{\text{Hys}}$ required just a modification specifying that a reset of the backoff stage (or {\tt STATE\_CW} in this case) is performed only when a packet is dropped or when the MAC queue empties.

To incorporate stickiness into the prototype we added a {\tt STATE} to the system that can be either {\tt STATE\_DETERMINISTIC} or {\tt STATE\_RANDOM} (related to the type of backoff being used), and a {\tt STICKINESS} counter that we reset each time we have a success and decrement when failing. When the counter gets to zero we enter the random state. Upon a successful transmission we unconditionally enter the deterministic state and reset the stickiness counter to {\tt DEFAULT\_STICKINESS}.

For the Schedule Reset mechanism, we added a bitmap for monitoring the state of every single slot after a successful transmission. To index the current slot in the bitmap we used the hardware register {\tt SPR\_IFS\_BKOFFDELAY}, which counts how many slots have still to come before the next transmission. The value of the register is decremented once per idle slot.

To avoid spurious detection, instead of continuously checking the state of the channel, we rely on the execution of the {\tt rx\_plcp} handler, which is called each time a valid PLCP is detected. When this happens, we implicitly know that the backoff counter has been frozen at least $20\mu s$ ago, which is the time between the detection of the first short trailing sequence and the complete decoding of the PLCP signal data. In any case, the slot is marked as busy.

We have the option to keep filling the bitmap for up to {\tt BITMAP\_ROUND\_BUILD} consecutive successful transmissions ($\gamma$ in Section~\ref{schedReset}) before checking if the central slot in the bitmap is available. If that is the case, the node's current schedule is halved (a schedule \emph{halving}, following Algorithm~\ref{alg:schedRest}) and its stickiness is incremented to {\tt DEFAULT\_STICKINESS}$+1$.

We performed four experiments for each tested number of contenders, starting from 1 and up to 25. For every experiment each station establishes an iPerf~\cite{tirumala2005iperf} session and transmits saturated UDP traffic towards a central AP, which also functions as iPerf server for each flow. We used WiFi channel 14  in order to avoid interference from other networks. The aggregated throughput is derived from the iPerf logs, while the percentage of lost frames is reported by the firmware. This is known to be $(\text{tx}-\text{sx})/\text{tx}$; where tx are the number of transmission attempts, and sx are the number of acknowledged frames.

\begin{figure}[tb]
		\centering
		\includegraphics[width=\linewidth,angle=-90]{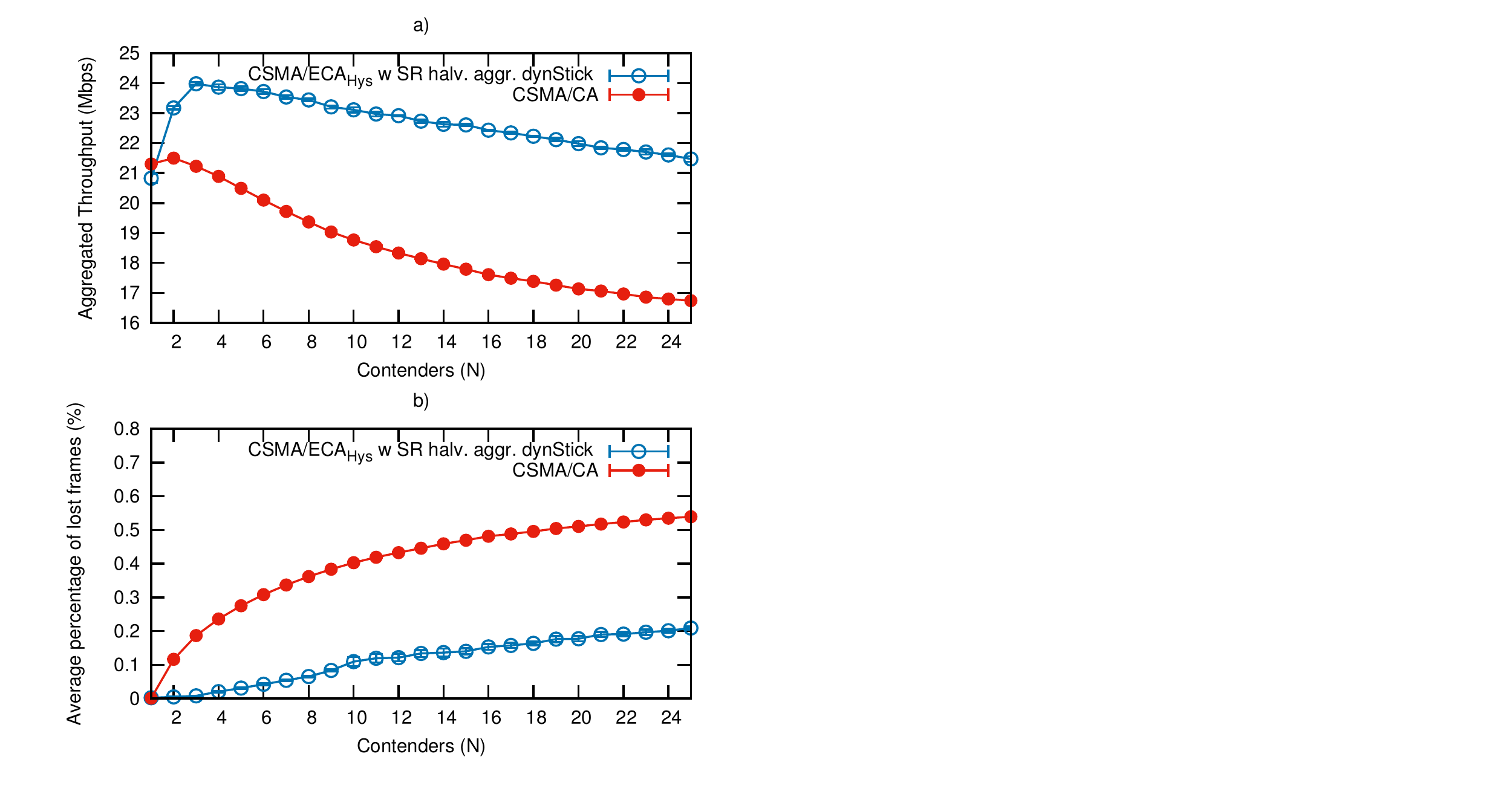}
		\caption{Implementation results: a) Average aggregated throughput for a saturated network. Real hardware implementation results (see Table~\ref{tab:testbed}); b) Average percentage of losses for a saturated network. Real hardware implementation results}
		\label{fig:implementationResults}
	\end{figure}

Figure~\ref{fig:implementationResults}a shows the average aggregated throughput, while Figure~\ref{fig:implementationResults}b presents the average percentage of lost frames. Details of the testbed are shown in Table~\ref{tab:testbed}. 

CSMA/ECA$_{\text{Hys}}$ has greater throughput due to its ability to avoid collisions more efficiently than CSMA/CA. Further, the Schedule Reset aggressiveness (see Section~\ref{aggr}) prevents CSMA/ECA$_{\text{Hys}}$ nodes from increasing too much the time between successful transmissions. 

Nevertheless, the implementation results reveal that there are other underlying factors that disrupt collision-free schedules. This is evidenced by the increasing percentage of lost frames followed by a throughput degradation in CSMA/ECA$_{\text{Hys}}$. The analysis of the factors that may disrupt collision-free schedules on real hardware implementations of CSMA/ECA$_{\text{Hys}}$ is left as future research.

	\begin{table}
		\centering
		\caption{PHY, MAC and other parameters for the real hardwre experiments}
		\label{tab:testbed}
		\begin{tabular}{|c|c|}
			\hline
			\multicolumn{2}{|c|}{{\bfseries PHY}}\\
			\hline
			{\bfseries Parameter} & {\bfseries Value}\\
			\hline
			PHY rate & 48~Mbps\\
			Empty slot & $9~\mu s$\\
			DIFS & $28~\mu s$\\
			SIFS & $10~\mu s$\\
			IEEE 802.11g WiFi channel & 14\\
			\hline
			\multicolumn{2}{|c|}{{\bfseries MAC}}\\
			\hline
			{\bfseries Parameter} & {\bfseries Value}\\
			\hline
			Maximum backoff stage ($m$) & 5\\
			Minium Contention Window ($CW_{\min}$) & 16\\
			Maximum retransmission attempts & 6\\
			Data payload (Bytes) & 1470\\
			Schedule Reset $\gamma$ & 1\\
			Schedule Reset mode & halving, dynStick\\
			Default stickiness & 1\\
			
			\hline
			\multicolumn{2}{|c|}{{\bfseries TESTBED}}\\
			\hline
			N & 25\\
			Distance between nodes and AP & 8~m\\
			Arrangement of nodes & Semicircle\\
			\hline
		\end{tabular}
	\end{table}

\section{Transitioning towards CSMA/ECA$_{\text{Hys+FS}}$}\label{ECAtoCA}


The current PHY/MAC enhancements considered by the HEW Task Group seek to achieve higher throughput and include improved coding schemes; full duplex radios, capable of receiving and transmitting at the same time; OFDMA and Multiple-User Multiple-Input Multiple-Output (MU-MIMO), allowing the transmission of different packets to/from multiple destinations at the same time; as well as dynamic channel bonding techniques~\cite{BorisChannelBonding}. Using CSMA/ECA$_{\text{Hys+FS}}$ alongside these features would provide enhanced performance by constructing collision-free schedules, thus substantially decre{black}asing the time spent recovering from collisions. 


Additionally, there are several features and scenarios still to be analysed for CSMA/ECA$_{\text{Hys+FS}}$ networks. Part of what is left for future work is summarised in the following:

\begin{itemize}
	\item The performance of CSMA/ECA$_{\text{Hys+FS}}$ WLANs in dense scenarios: transmissions from stations in the same or other networks may negatively affect the collision-free operation, particularly when not all devices in the CSMA/ECA$_{\text{Hys+FS}}$ WLAN are able to listen such transmissions. As a consequence, not all stations will pause the backoff countdown accordingly, resulting in large slot drifts that disrupt any collision-free schedule, approximating CSMA/ECA$_{\text{Hys+FS}}$ performance to CSMA/CA's. 

	
	\item Traffic differentiation: although priorities using different contention windows in CSMA/ECA$_{\text{Hys+FS}}$ proved to outperform CSMA/CA, other Enhanced Distributed Channel Access (EDCA) mechanisms like the Arbitration Inter-Frame Spacing (AIFS) would not work~\cite{jaumeTD}, whereas the following EDCA mechanisms would:
	\begin{itemize}
		\item Transmission Opportunity (TXOP): stations with an increased TXOP are able to transmit more packets. APs in CSMA/ECA$_{\text{Hys+FS}}$ can use a big TXOP in order to transmit more than users.
		\item Multiple Queues:~\cite{jaumeTD} provides performance metrics for two traffic categories. In order to transition to CSMA/ECA$_{\text{Hys+FS}}$ a total of four access categories (AC) should be implemented, as in EDCA. Priorities to the multiple queues can be granted through different minimum and maximum contention windows.
	\end{itemize}
	CSMA/ECA$_{\text{Hys+FS}}$ with multiple access categories is expected to provide better traffic differentiation in WLANs, mainly due to the elimination of collisions by using a deterministic backoff after each access category's successful transmission, as well as providing a minimum amount of transmission opportunities to each AC.
	\item Coexistence with EDCA stations in the same WLAN.
\end{itemize}

\section{Conclusions}\label{conclusions}
CSMA/ECA$_{\text{Hys+FS}}$ is able to construct a collision-free schedule with many contenders. Taking advantage of this condition, the cumulative throughput experienced by CSMA/ECA$_{\text{Hys+FS}}$ nodes goes beyond the achievable by CSMA/CA for any number of nodes. All of these while preserving throughput fairness. Further, as non-saturated CSMA/ECA$_{\text{Hys+FS}}$ networks get crowded the cumulative throughput keeps increasing up to the saturation point, contrary to the throughput degradation seen in CSMA/CA networks.

As CSMA/ECA$_{\text{Hys+FS}}$ is thought to be the next MAC for WLANs, coexistence with CSMA/CA nodes is paramount. Results show that coexistence is not an issue for any proportion of CSMA/ECA$_{\text{Hys+FS}}$/CSMA/CA users. Moreover, when the network is composed with a majority of CSMA/ECA$_{\text{Hys+FS}}$ nodes, the cumulative throughput increases from CSMA/CA's.

To top it all, the real hardware implementation of CSMA/ECA$_{\text{Hys}}$ clearly outperforms CSMA/CA. In itself, the prototype represents a strong indicator that the proposed enhanced collision avoidance could be considered as a viable replacement or evolution of the current MAC for WiFi.


\bibliographystyle{IEEEtran}
\bibliography{ref}

\end{document}